\documentclass[useAMS,usenatbib,letterpaper]{mn2e}
\usepackage[pdftex]{graphicx}
\usepackage{amsmath}
\makeatletter

\def\compoundrel#1\over#2{\mathpalette\compoundreL{{#1}\over{#2}}}
\def\compoundreL#1#2{\compoundREL#1#2}
\def\compoundREL#1#2\over#3{\mathrel
      {\vcenter{\hbox{$\m@th\buildrel{#1#2}\over{#1#3}$}}}}
\usepackage{subfigure}
\def\lesssim{\mathrel{\hbox{\rlap{\hbox{\lower4pt\hbox{$\sim$}}}\hbox{$<$}}}}

\title[AGN jet transverse asymmetries]{Signatures of large-scale magnetic fields in AGN jets: transverse asymmetries}
\author[E. Clausen-Brown, M. Lyutikov, P. Kharb ]{E. Clausen-Brown$^{1}$\thanks{browner@purdue.edu}, M. Lyutikov$^{1}$, P. Kharb$^{2}$\\
$^{1}$Department of Physics, Purdue University, 525 Northwestern Avenue, West Lafayette, IN 47907-2036, USA\\
$^{2}$Department of Physics, Rochester Institute of Technology, Rochester, USA}
\begin{document}

\date{Accepted}

\pagerange{\pageref{firstpage}--\pageref{lastpage}} \pubyear{2011}

\maketitle

\label{firstpage}

\begin{abstract}
We investigate the emission properties that a large-scale helical magnetic field imprints on AGN jet synchrotron radiation.  A cylindrically symmetric relativistic jet and large-scale helical magnetic field produce significant asymmetrical features in transverse profiles of fractional linear polarization, intensity, Faraday rotation, and spectral index.  The asymmetrical features of these transverse profiles correlate with one another in ways specified by the handedness of the helical field, the jet viewing angle ($\theta_{ob}$), and the bulk Lorentz factor of the flow ($\Gamma$).  Thus, these correlations may be used to determine the structure of the magnetic field in the jet.  In the case of radio galaxies ($\theta_{ob} \sim 1$ radian) and a subclass of blazars with particularly small viewing angles  ($\theta_{ob} \ll 1/\Gamma$), we find an edge-brightened intensity profile that is similar to that observed in the radio galaxy M87.  We present observations of the AGNs 3C 78 and NRAO 140 that display the type of transverse asymmetries that may be produced by large-scale helical magnetic fields.
\end{abstract}

\section{Introduction}
\label{section:intro}
Large-scale magnetic fields are thought to play a central role in launching relativistic jets in active galactic nuclei, or AGN \citep{Blandford:1977,Blandford:1982,Lovelace:1987}.  Following the initial production of the jets, the large-scale magnetic field naturally develops into a helical configuration in which the toroidal component of the field collimates the outflow into a narrow jet via magnetic hoop stresses \citep{Benford:1978,Chan:1980}.  While the importance of large-scale magnetic fields in the initial production of jets is rarely disputed, it is not known how far from the nucleus the ordered large-scale component of the field continues to remain dominant over the tangled component.  Current-driven instabilities, of which the $m = 1$ kink mode is most important, are prime candidates for tangling or mixing the large-scale helical field when it is toroidally dominated \citep{Begelman:1998,Giannios:2006}.  The kink mode instability may imply that the field becomes tangled soon after the large-scale field has launched, accelerated, and collimated the jet \citep[e.g.,][]{Marscher:2008}, or it may imply that only tangled fields \citep{Heinz:2000} or only poloidal fields \citep{Spruit:1997} play a significant role in astrophysical jets.  Alternatively, the growth of the kink mode may be stabilized through various effects such as gradual shear or an external wind \citep[][and references therein]{Hardee:2004,Mckinney:2009}.  Clearly, observations are needed to further illuminate whether parsec scale AGN jets contain large-scale helical magnetic fields.

The most direct method of observing the existence and geometry of large-scale magnetic fields, very long baseline interferometry (VLBI) polarimetry, yields largely ambiguous results.  Parsec scale radio jet synchrotron emission is often highly polarized, implying that the jet magnetic field is anisotropic but not necessarily large-scale \citep{Pacholczyk:1970}.  The electric vector position angle (EVPA) of a synchrotron emitting element is perpendicular to the element's magnetic field direction projected onto the sky.  However, due to relativistic aberration in AGN jets the EVPA is not necessarily perpendicular to the observer frame jet magnetic field  \citep{Lyutikov:2005}.  Statistically, observed EVPAs have a bimodal distribution: most are either aligned with the local jet axis or perpendicular to it \citep{Cawthorne:1993,Marscher:2002,Lister:2005,Kharb:2008}.  Two classes of models explain these data: (i) models assuming a helical field and a cylindrical jet \citep{Lyutikov:2005} and (ii) shock and velocity shear models which assume that shock compression along the jet axis produces EVPAs parallel to the jet axis and that velocity shear explains EVPAs perpendicular to the jet axis \citep{Laing:1980,Hughes:1989a,Hughes:1989b,Hughes:1991,Attridge:1999,Kharb:2005}.  Both models explain the presence of linear polarization, which would be absent if the magnetic field were tangled and isotropic.

To distinguish between these two explanations of VLBI polarimetry data and more generally test the assumption of parsec scale helical magnetic fields, several researchers have calculated how intensity and fractional polarization change across a jet in the transverse direction (i.e. profiles).  Laing et al. \citeyearpar{Laing:1981,Laing:2006} compared the polarization and intensity profiles of kpc jets to a variety of helical field models and tangled anisotropic field models and found the predicted asymmetries of helical models to be inconsistent with observed symmetrical profiles.  However, \cite{Papageorgiou:2006} found a number of similarities between the intensity and polarization profiles due to helical fields calculated in \cite{Laing:1981} and AGN jet observations.  Similarly, \cite{Aloy:2000} carried out three-dimensional hydrodynamnic jet simulations that found asymmetries similar to those discussed in this paper in profiles of intensity and polarization.  \cite{Lyutikov:2005} studied the polarization profiles and other polarization properties of relativistic jets with helical fields, finding that the presence of helical fields explains both the bimodal distribution of EVPAs for unresolved jets and the abrupt $90^{\circ}$ flips in the EVPA for jets with small viewing angles (i.e. blazars).  

Recently, observed gradients in Faraday rotation measure ($RM$) have received much attention.  Following the suggestion that helical fields cause $RM$ gradients across jets \citep{Laing:1981,Blandford:1993}, observers have firmly established such behavior in the famous quasar 3C 273 \citep{Asada:2002,Zavala:2005} and have also found gradients in other parsec scale AGN jets \citep{Asada:2008b,Gabuzda:2004,Gabuzda:2008,Kharb:2009,O'Sullivan:2009,Croke:2010}.  These observations have prompted a number of efforts to fit the observed $RM$ profiles to simple analytic models of AGN \citep{Contopoulos:2009,Konigl:2010} as well as numerical simulations of AGN \citep{Broderick:2010}.

To investigate the question of large-scale helical fields, we calculate transverse profiles using the polarized synchrotron absorption and emission coefficients, retaining the coefficients' dependence on the angle between the jet frame line of sight and magnetic field.  These calculated profiles show that a jet having an axially symmetric structure and helical magnetic field nevertheless display asymmetric profiles in intensity, Faraday rotation, fractional polarization, and spectral index due to the large-scale helical structure of the magnetic field.  In this work we do not discuss the observations or modeling of EVPA swings of bright emission features in parsec-scale AGN jets which may help further our understanding of AGN jet magnetic fields \citep{Marscher:2008,Abdo:2010}. This paper is structured as follows.  In \S\ref{section:model}, we discuss the assumptions made regarding the structure and geometry of the relativistic jet.  Optically thin intensity profiles are calculated and compared to observations in \S\ref{section:intensity}.  Parsec scale $RM$ calculations and observations in \S\ref{section:RM} are followed by unconvolved and convolved calculations of polarization profiles and polarization observations in \S\ref{section:pi}.  Spectral index profiles are calculated in \S\ref{section:alpha}.  Finally, in \S\ref{section:conclusion} we summarize our work and discuss how asymmetrical features in each of the aforementioned profiles should correlate with one another.  As an example of how to employ the predicted correlations of our helical model, in \S\ref{section:conclusion} we also briefly discuss possible correlations seen amongst the different transverse profiles of the radio jets 3C 78 and NRAO 140.
\section{\normalsize A Simple Model}
\label{section:model}
\subsection{Magnetic Field Structure}
\label{section:Mag}
A crucial assumption we make regarding the structure of the helical magnetic field is that the jet frame azimuthal and axial fields are of comparable magnitude, $B_{\phi}'/B_z' \sim 1$ (all primed quantities refer to the rest frame of the jet).  As explained in \S\ref{section:geo}, if this assumption is false and the jet frame magnetic field probed by parsec scale VLBI observations is dominated by either the azimuthal or axial field, then the transverse profiles described here would not exhibit the asymmetries that are the subject of this paper (except for the $RM$ profiles).  Ideal magnetohydrodynamic (MHD) models of jets launched from accretion disks predict that, in the conical region of the jet, the observer frame azimuthal to axial field ratio at a distance $z$ above the disk is $B_{\phi}/B_z \approx R_j(z) \Omega/c$, where $\Omega$ is the angular frequency of the magnetic field line's footpoint anchored in the accretion disk at a radius $r_0$ from the central black hole, and $R_j$ is the cylindrical radius of the jet at a distance $z$ above the accretion disk \citep[e.g.,][]{Appl:1993}.  Assuming the field line footpoint is located at $r_0=\eta GM/c^2$ and is in Keplerian orbit around a black hole of mass $M$, we estimate the following for the jet frame ratio for typical VLBI AGN jet parameters:
\begin{gather}
\frac{B'_{\phi}}{B_z'} \sim 6\left(\frac{\eta}{10}\right)^{-3/2} \left(\frac{R_j}{0.1\mbox{pc}}\right) \left(\frac{\Gamma}{10}\right)^{-1} \left(\frac{M}{10^9M_{\sun}}\right)^{-1}.
\label{eqn:ratio}
\end{gather}
This simple estimate for $B_{\phi}'/B_z'$ is sensitive to the jet production location, $r_0$---parameterized by $\eta$ in equation (\ref{eqn:ratio})---which has been estimated to be anywhere from $\eta \sim$ few to several tens depending on the spin of the black hole and other properties of the AGN system \citep{Meier:2001,Vlahakis:2004}.  If this wide range of values for $\eta$ is realized in nature, then it is possible that a modest fraction of parsec and sub-parsec scale jets probed by VLBI experiments have magnetic fields where $B_{\phi}'/B_z'\sim 1$, producing the asymmetric transverse profiles discussed in this paper.

Alternatively, $B_{\phi}'/B_z'$ may not depend on conditions at the base of jets as it does in equation (\ref{eqn:ratio}).  If ultra-relativistic AGN jets are magnetically dominated on parsec scales \citep{Blandford:2002,Vlahakis:2004} and conditions on the boundary of the jet do not change too rapidly, the jet may relax to a force-free structure, ${\bf \nabla} \times \vec{B}=k\vec{B}$ \citep{Choudhuri:1986}.  Force-free fields have the property that $B_{\phi}'/B_z' \sim 1$, implying that significant transverse asymmetries are expected in VLBI profiles.  Two simple force-free configurations to which a cylindrical jet could relax are (i) a diffuse pinch in which $k$ depends on position \citep[e.g.,][]{Lynden-Bell:1996} and (ii) the option used in this paper: a reverse field pinch where $k=$ constant \citep{Lundquist:1950,Choudhuri:1986}.  The minimum energy configuration of a helicity conserving magnetic field corresponds to a force-free field in which $k=$ constant \citep{Woltier:1958}.  Thus, if the jet frame field relaxes via magnetic dissipation while approximately conserving helicity, the field may find the minimum energy state \citep{Taylor:1974}, which is the reverse field pinch for a cylindrical geometry:
\begin{equation}
\vec{B}'(\rho,\phi,z)=B_0\left[0,J_1(k\rho),J_0(k\rho) \right],
\label{eqn:B}
\end{equation}
where $J_{1,2}$ are Bessel functions of the first kind, and the cylindrical coordinate, $\rho$, is normalized so $\rho=1$ at the boundary of the jet.  As can be seen in figure \ref{fig:field}, we set $k\cong 2.405$, to ensure $B_z=0$ at $\rho=1$, such that there are no reversals of the axial field in the emission region, or jet spine.  For the purposes of calculating the $RM$, we locate the Faraday rotating region outside the emission region between cylindrical radii $\rho=1$ and $\rho=1.6$.  The decision to locate the Faraday rotating region outside the emission region is motivated by a variety of observations described in \S\ref{section:RM}.
\subsection{Jet Geometry and Emission Properties}
\label{section:geo}
We assume the jet is a steady cylindrical flow of radius $R_j$ with a bulk Lorentz factor of $\Gamma=(1-\beta^2)^{-1/2}=10$, where $\vec{\beta}=\vec{v}/c=\beta (0,0,1)$ is the jet speed in units of the speed of light expressed in cylindrical coordinates $(\rho,\phi,z)$ centered on the jet axis.  The observer is located in the $\phi=0$ (or $x$-$z$) plane such that observed photons move along the unit vector $\vec{n}=\left(\sin{\theta_{ob}},0,\cos{\theta_{ob}}\right)$, where $\theta_{ob}$ is the angle between the photon propagation vector, $\vec{n}$, and the jet propagation direction, $\hat{z}$.

The cylindrical jet approximation used here breaks down when the jet is viewed with $\theta_{ob} \leq \theta_j$, where $\theta_j$ is the half opening angle of a more realistic conical jet.  Thus, comparing our model to blazars (AGN where $\theta_{ob} \lesssim 1/\Gamma$) might seem questionable except for the simple trend in VLBI radio jet surveys such as the MOJAVE sample \citep{Lister:2009}: a significant majority of MOJAVE sources have a core-jet structure suggesting that, typically, $\theta_{ob}>\theta_j$.  A jet viewed with $\theta_{ob} \leq \theta_j$ would not have a simple core-jet morphology.  When viewed at such small angles, the jet would have features moving away from its core in all directions on the sky, leaving a core without a distinct jet morphology.  As the VLBI images of most of the sources in MOJAVE reveal a clear core-jet morphology, we can conclude that they are viewed with $\theta_{ob}> \theta_j$.
\begin{figure}
\centering
\subfigure[][\label{fig:helix}] {\includegraphics[width=2in]{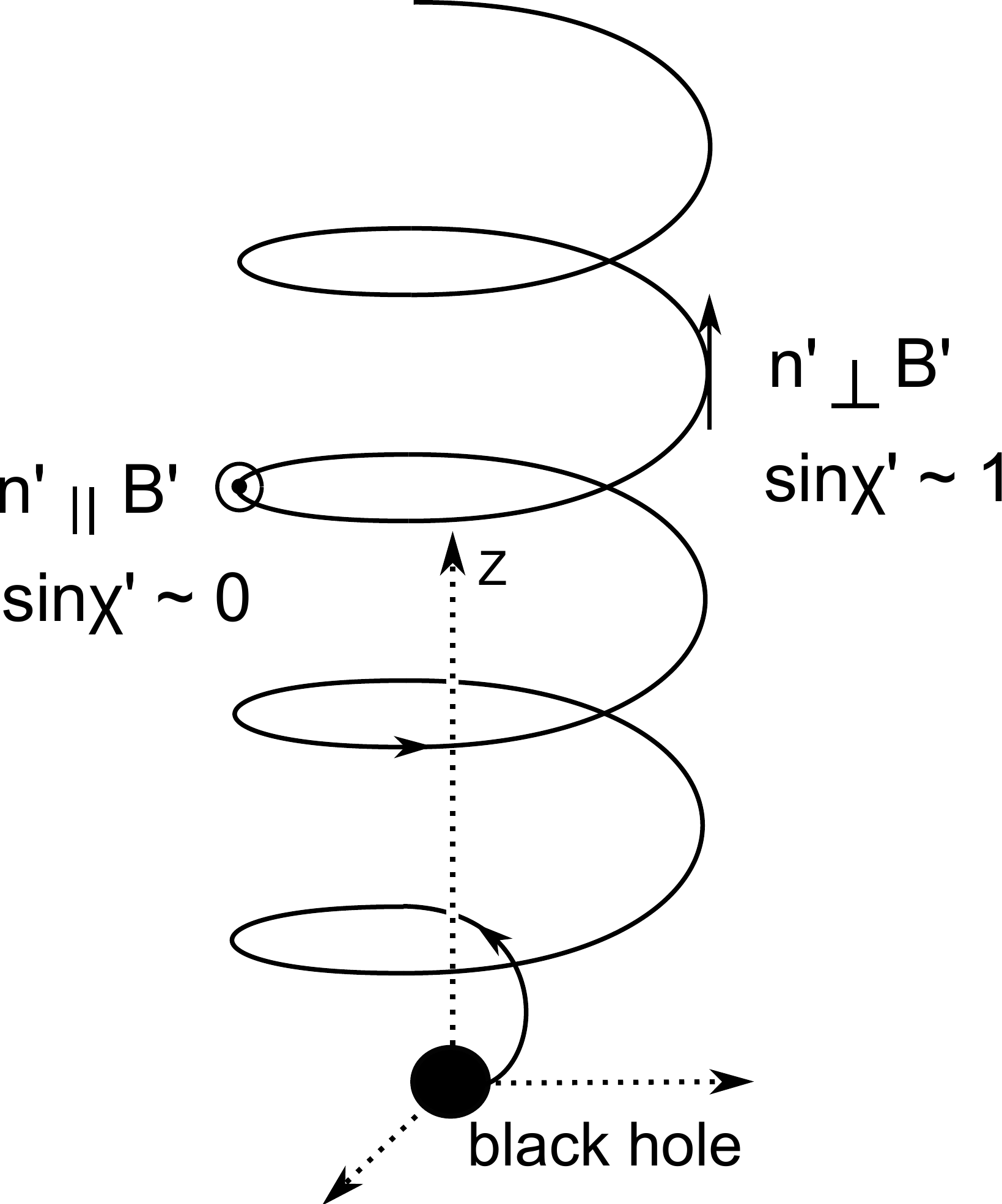}}
\subfigure[][\label{fig:field}] {\includegraphics[width=3in]{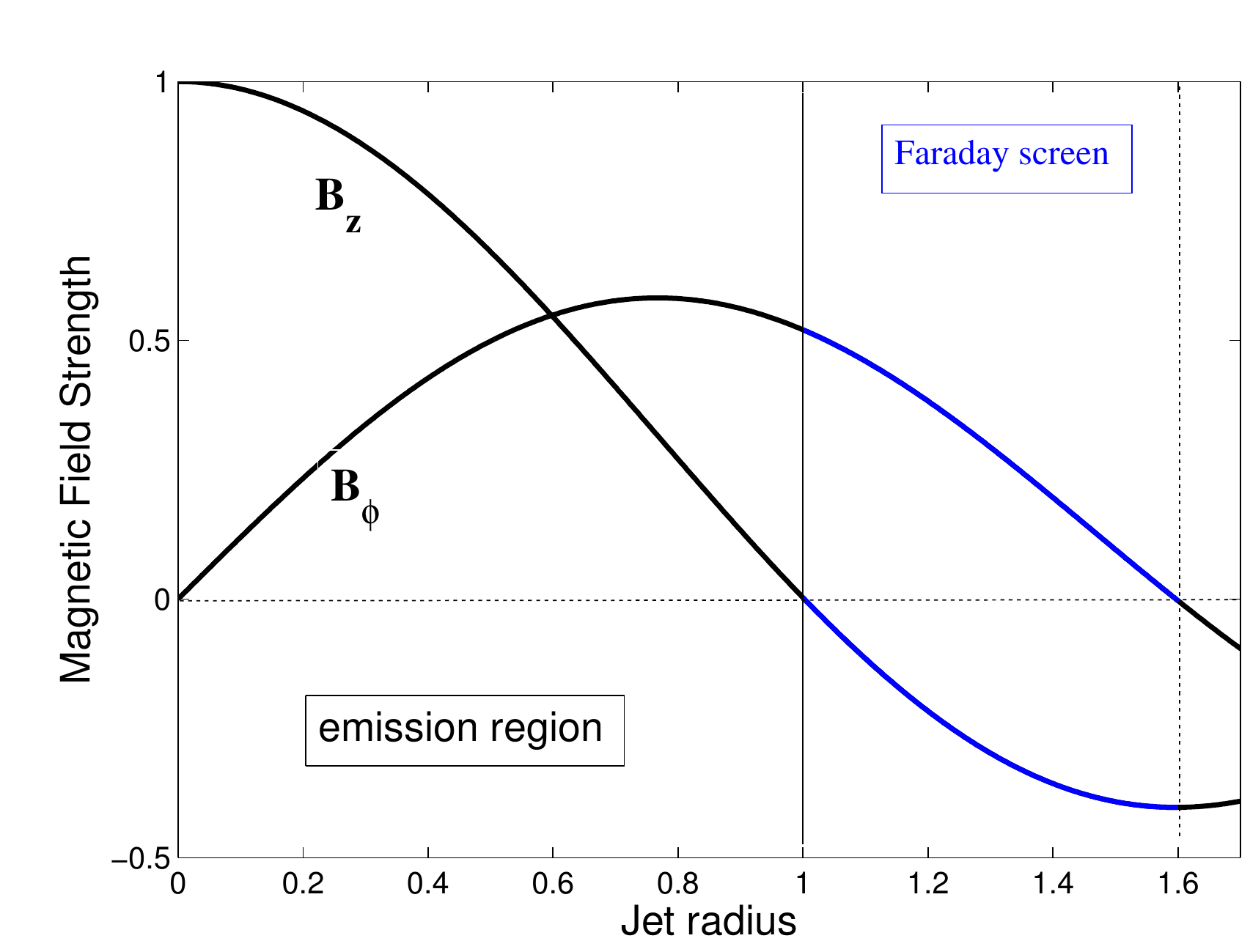}}
\caption{(a) This schematic shows one magnetic field line in the jet comoving frame from the observer's perspective where $\theta_{ob}' < \pi/2$ and $B_{\phi}'/B_z'\sim 1$.  On the right side of the jet $\sin{\chi'} \sim 1$, that is, the magnetic field is approximately perpendicular to the observer's line of sight.  On the left side $\sin{\chi'} \sim 0$, meaning that the magnetic field is pointing almost directly at the viewer.  This asymmetry manifests itself in VLBI profiles of intensity, polarization, spectral index, and $RM$.  Hereafter, the terms ``left" and ``right" will be referencing the sides of a jet viewed with $\theta_{ob}'<\pi/2$ that is filled with a right-handed helical magnetic field as shown here.  (b) This is a plot of equation (\ref{eqn:B}) as a function of cylindrical radius, $\rho$, which is the magnetic field configuration used in this work.  Note that the pitch angle of the magnetic field configuration varies with cylindrical radius such that the field is purely axial on the jet axis and purely azimuthal on the jet boundary.} 
\label{fig:schematic}
\end{figure}

We assume the jet synchrotron emission is produced by a power-law distribution of relativistic electrons, $dn'=K_e' E'^{-p}dE'$.  The polarized emission coefficients in the plasma rest frame for a power-law distribution of electrons can be expressed as (leaving out the primes)
\begin{align}
j_{\nu}^{(i)}=&c_1^{(i)}K_e\left|B\sin{\chi}\right|^{\frac{p+1}{2}}\nu^{-(p-1)/2} \notag \\
c_1^{(i)}=&\frac{\sqrt{3}e^3}{32\pi m_e c^2}\left(\frac{3e}{2\pi m_e^3c^5}\right)^{(p-1)/2}\tilde{\Gamma}\left(\frac{3p-1}{12}\right)\notag \\
& \times\tilde{\Gamma}\left(\frac{3p+7}{12}\right)\left(\frac{p+7/3}{p+1} \pm 1\right),
\label{eqn:emission}
\end{align}
and the absorption coefficients are
\begin{align}
\kappa_{\nu}^{(i)}=&c_2^{(i)}K_e\left|B\sin{\chi}\right|^{(p+2)/2}\nu^{-(p+4)/2} \notag\\
c_2^{(i)}=& \frac{\sqrt{3}e^3}{32 \pi m_e}\left(\frac{3e}{2\pi m_e^3c^5}\right)^{p/2}\left(p+\frac{10}{3}\right)\tilde{\Gamma}\left(\frac{3p+2}{12}\right) \notag\\
&\times \tilde{\Gamma}\left(\frac{3p+10}{12}\right)\left(1\pm \frac{p+2}{p+10/3}\right),
\label{eqn:absorption}
\end{align}
where $\chi$ is the angle between the magnetic field and line of sight (i.e. $B\cos{\chi}=\vec{B} \cdot \vec{n}$), $m_e$ is the electron mass, $e$ is the elementary charge, $\nu$ is the frequency of the electromagnetic wave, and $\tilde{\Gamma}(x)$ is the gamma function of argument $x$ \citep{Pacholczyk:1970}.  The $i=1$ ($i=2$) state refers to the value of the coefficients with the upper sign (lower sign) in equations (\ref{eqn:emission}) and (\ref{eqn:absorption}).  These coefficients correspond to orthogonal radiation polarization states for which the photon electric vector is perpendicular (parallel) to the component of the local magnetic field projected onto the sky (i.e. perpendicular to the photon propagation direction).

The asymmetry highlighted in this paper is that the left and right sides of the jet axis projected onto the sky have different synchrotron emission properties \citep[cf.][]{Aloy:2000}.  The cause of this asymmetry is most clearly illustrated by the representation of a single magnetic field line viewed in the jet frame in figure \ref{fig:helix}, where the left side clearly appears different from the right side of the helix.  (Hereafter, when referencing the ``left" or ``right" side of the jet, we are referring to the emission region of a jet with a helical field that is a right-handed helix for which $\theta_{ob}'<\pi/2$ as shown in figure \ref{fig:helix}.)  More precisely, $\left\langle \left|\sin{\chi'}\right| \right\rangle_{left}$ on the left side of the jet is less than $\left\langle \left|\sin{\chi'}\right| \right\rangle_{right}$ on the right side of the jet, where $\chi'$ is the angle between the jet frame line of sight, $\vec{n}'$, and magnetic field, $\vec{B}'$.  The $\left\langle \right\rangle$ symbols refer to an emission weighted average along the line of sight intersecting the jet. As discussed in \S\ref{section:Mag}, there is only a significant asymmetry in $\left\langle \left|\sin{\chi'}\right| \right\rangle$ if $B_{\phi}'/B_z'\sim1$; otherwise, if the field is azimuthally dominated or axially dominated, then $\left\langle \left|\sin{\chi'}\right| \right\rangle_{left} \sim \left\langle \left|\sin{\chi'}\right| \right\rangle_{right}$ and no transverse asymmetry will be observed.  This asymmetry in $\left\langle \left|\sin{\chi'}\right| \right\rangle$ can clearly cause asymmetries in the jet's synchrotron properties as both the synchrotron emission and absorption coefficients depend on $\left|\sin{\chi'}\right|$ as seen in equations (\ref{eqn:emission}) and (\ref{eqn:absorption}).  

Note that the asymmetry in $\left\langle \left|\sin{\chi'}\right| \right\rangle$ is not intrinsic, but instead depends on the jet frame viewing angle $\theta_{ob}'$, which is connected to $\Gamma$ and $\theta_{ob}$ via the relativistic aberration effect.  For $\theta_{ob}'=\pi/2$ there is no asymmetry because $\left\langle \left|\sin{\chi'}\right| \right\rangle_{left} = \left\langle \left|\sin{\chi'}\right| \right\rangle_{right}$.  The calculated profiles are asymmetric when the jet frame viewing angle is $\theta_{ob}'=\pi/2\pm\xi$, where the constant $\xi$ is between $0$ and $\pi/2$.  Profiles originating from a jet with $\theta_{ob}'=\pi/2+\xi$ or a jet with $\theta_{ob}'=\pi/2-\xi$ are reflected versions of the other, where the reflective symmetry axis is the z-axis projected onto the sky.  For example, if an AGN jet with $\theta_{ob}'=\pi/2+\xi$ and a positively skewed profile has its z-axis rotated so that $\theta_{ob}'=\pi/2-\xi$, then the rotated AGN jet's new profile will be negatively skewed (i.e. reflected about the z-axis projected onto the sky).  In a relativistic flow where $\Gamma \gg 1$, no asymmetry is expected for the observer frame viewing angle $\theta_{ob}\approx1/\Gamma$, corresponding to $\theta_{ob}'\approx\pi/2$.  Furthermore, the observer frame viewing angles $\theta_{ob}=N/\Gamma$ or $1/(N\Gamma)$, where $N\ll\Gamma$, correspond to $\theta_{ob}'\approx \pi/2\pm\xi$.  Consequently, AGN jets viewed with $\theta_{ob}=N/\Gamma$ produce approximately the same profile as $\theta_{ob}=1/(N\Gamma)$, except with the profile being reflected about the jet axis projected onto the sky.
\section{Intensity profiles}
\label{section:intensity}
In a large-scale magnetic field, optically thin synchrotron emission is not isotropic because it depends on $\sin{\chi'}$.  Since the synchrotron emissivity is an increasing function of $\sin{\chi'}$, more power is emitted toward observers whose line of sight is perpendicular to the magnetic field.  Thus, for a helical magnetic field as seen in figure \ref{fig:helix}, more synchrotron intensity will be observed on the right side of the jet ($a>0$) where $\sin{\chi'}\sim1$ than on the left side where $\sin{\chi'}\sim0$.  We demonstrate this behavior by integrating over the jet-frame emission coefficient defined in equation (\ref{eqn:emission}) for a power law distribution of electrons:
\begin{gather}
I_{\nu} \propto \int^{S'}_{0}{j_{\nu'}'}ds',
\label{eqn:Int}
\end{gather}
where $S$ is the total path length through the jet for a given line of sight.

\begin{figure}
\centering
\includegraphics[width=2.5in]{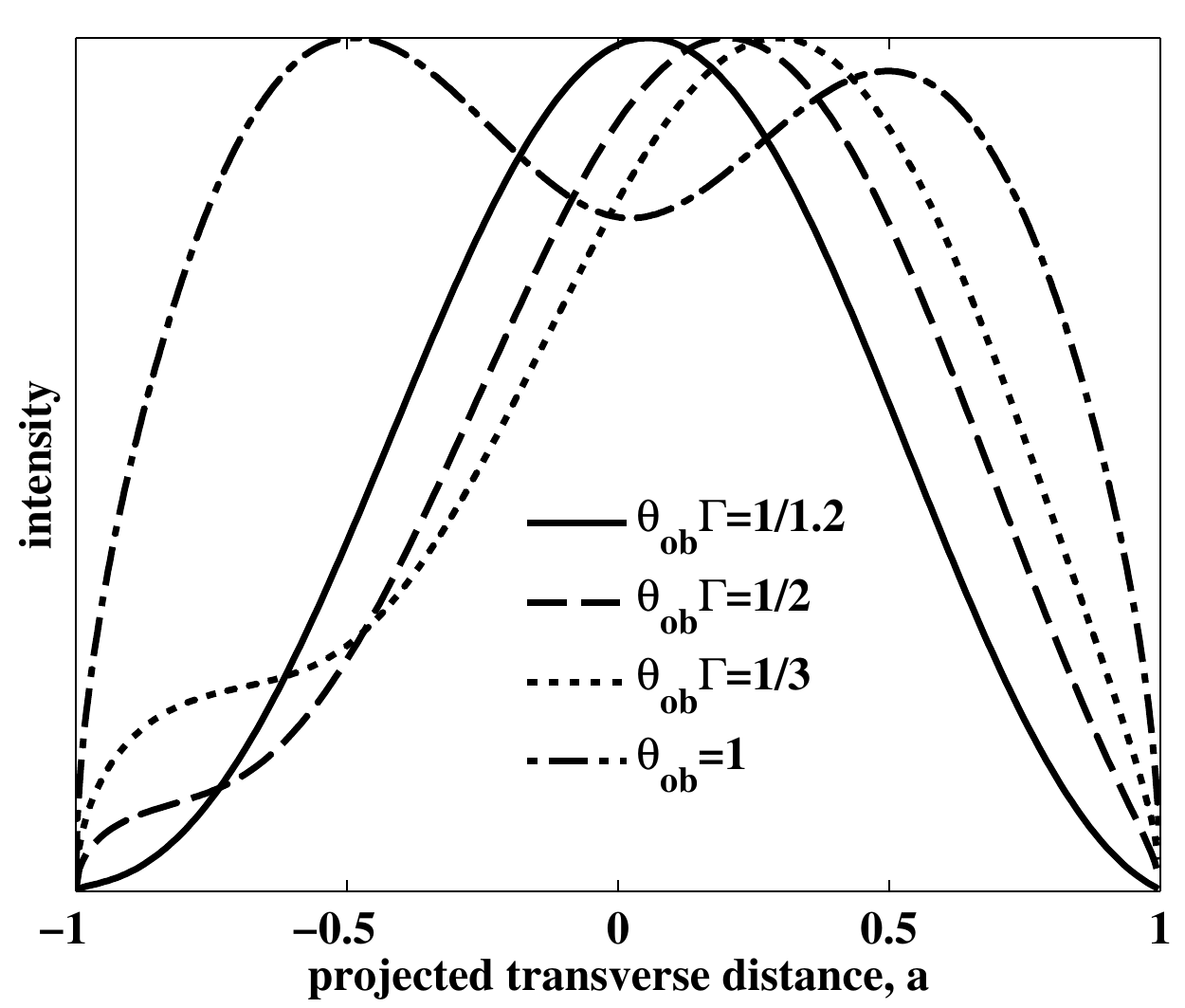}
\caption{This plot depicts theoretical intensity curves calculated as described in section \ref{section:intensity}, for observer frame (jet frame) angles of $\theta_{ob}\Gamma=1/1.2$ (74$^{\circ}$), $\theta_{ob}\Gamma=1/2$ (53$^{\circ}$), $\theta_{ob}\Gamma=1/3$ (37$^{\circ}$), $\theta_{ob}=1$ (170$^{\circ}$), where $\Gamma=10$. Except for the $\theta_{ob}=1$ curve, all of the curves are calculated using $\theta_{ob}\Gamma<1$, so that the jet frame viewing angle is $\theta_{ob}'< \pi/2$ and the jet is viewed as shown in figure \ref{fig:helix}.  Note that the intensity peaks occur on the right side of the jet ($a>0$) where, as shown in figure \ref{fig:helix}, the angle between the line of sight and the B-field, $\chi'$, is closer to $\sim \pi/2$ and therefore, according to equation (\ref{eqn:emission}), the emission is greater.  The exception to this trend is the $\theta_{ob}=1$ curve which represents jets that are close to being in the plane of the sky.  Radio galaxies with relativistic flows behave differently because the rest frame viewing angle is almost $\theta_{ob}'\sim \pi$.  } \label{fig:theory_I}
\end{figure}
\begin{figure}
\centering
\subfigure[ ][\label{fig:int3C78}]{\includegraphics[width=3in]{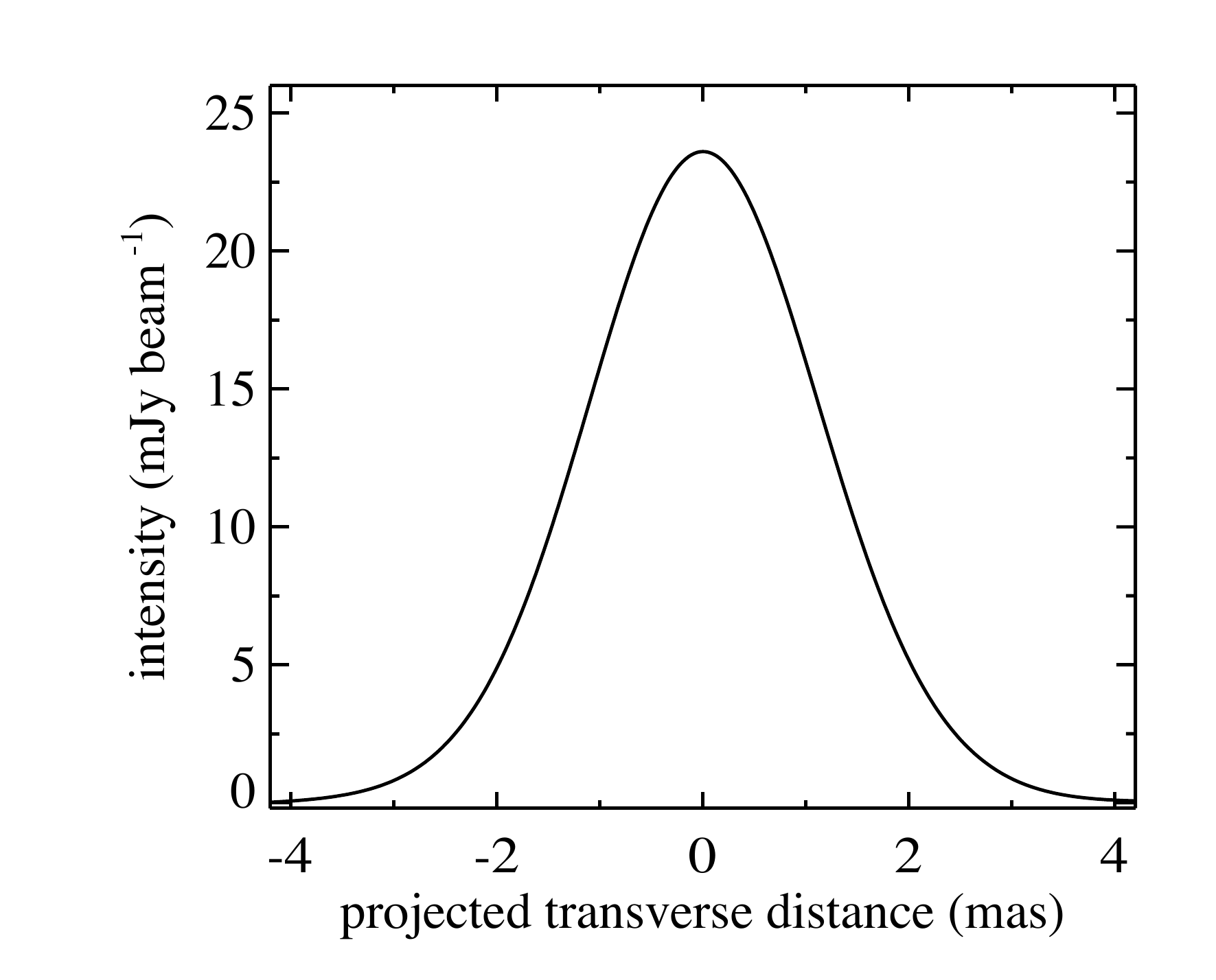}}
\subfigure[][\label{fig:intNRAO140}]{\includegraphics[width=3in]{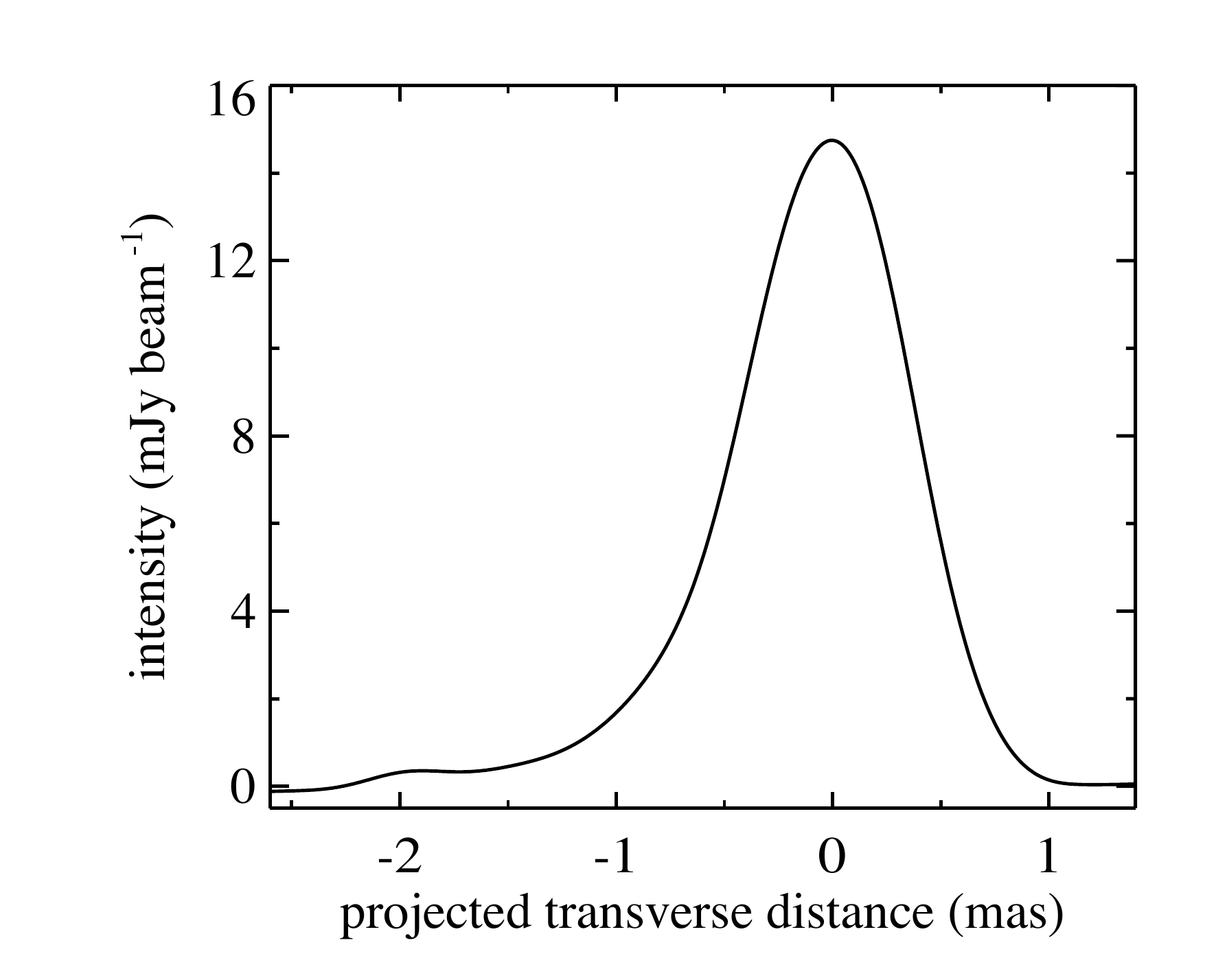}}
\caption{(a)  This observed $5$ GHz intensity profile of 3C 78, with a beam-size (FWHM) of $2.5$ mas $\times$ 2.5 mas, displays none of the predicted asymmetry.  This could be because predicted asymmetrical features are below the sensitivity of VLBA.  (b)  This observed $15$ GHz intensity profile of the quasar NRAO 140 (with a beam-size of $0.9$ mas $\times$ 0.6 mas) does exhibit a small asymmetry, a longer tail on the left, which is consistent with our theoretical profiles.}
\label{fig:int_obs}
\end{figure}
The results of these calculations with arbitrary normalization can be seen in figure \ref{fig:theory_I}.  Doppler beaming is not important for the shape of these profiles as it only affects their normalization.  As expected, the profiles go to zero at the edges of the jet and peak in the projected central region because lines of sight are longer through the central regions of the jet than on the sides.  The anisotropic synchrotron emission in the presence of the helical field is responsible for the off-center peaks and most importantly for the skewness of the profiles.  Thus, the primary observable effect helical fields have on intensity profiles is to skew them such that the longer tail of the profile is on the side of the jet where the magnetic field is approximately parallel to $\vec{n}'$.  An important transition occurs in the skewness of our profiles at $\theta_{ob}=1/\Gamma$.  Given the handedness of the helical field in figure \ref{fig:helix}, a jet viewed with $\theta_{ob}<1/\Gamma$ has intensity profiles with a negative skew (i.e. its long tail is to the left), while the same jet viewed with $\theta_{ob}>1/\Gamma$ has a positive skew.  The skewness is most clearly manifested for $\theta_{ob}<1/2\Gamma$ and $\theta_{ob}>2/\Gamma$; otherwise the jet's intensity profile will be approximately symmetric.  

As the $\theta_{ob}\sim1$ case in figure \ref{fig:theory_I} shows, the skewed profiles become double humped for either radio galaxies, $\theta_{ob} \sim 1$, or for blazars with $\theta_{ob} \ll 1/\Gamma$ since in both cases the rest frame viewing angles are $\theta_{ob}\sim \pi$ or $0$ respectively.  This edge-brightening occurs because for such jet viewing angles the jet frame line of sight is almost parallel to the jet axis.  Therefore, lines of sight crossing through the center of the jet where the axial field dominates will be parallel to the magnetic field, and little synchrotron radiation will be emitted towards the observer.  Closer to the edges of the jet the magnetic field is predominately toroidal and perpendicular to the line of sight, so more radiation is emitted toward the observer.  Thus, the presence of edge-brightening in jets such as M87 \citep{Reid:1989} and 3C 345 \citep{Unwin:1992} may be explained by jet viewing angles with $\theta_{ob} \gg$ or $\ll 1/\Gamma$.

Observed intensity profiles of the radio galaxy 3C 78 and blazar NRAO 140\footnote{NRAO 140 data obtained from the MOJAVE database at http://www.physics.purdue.edu/MOJAVE/allsources.shtml} are shown in figures \ref{fig:int3C78} and \ref{fig:intNRAO140} respectively.  3C 78 is a radio galaxy with $\theta_{ob}\sim1$ and a mildly relativistic parsec scale jet ($\beta\leq 0.52$) as indicated by its parsec scale jet to counterjet surface brightness ratio \citep{Kharb:2009}.  The symmetry of 3C 78's profile may be explained by the fact that the jet frame line of sight is perpendicular to the jet axis, or it is possible that the finite beamwidth smears out any small skewness the true intensity profile contains.  NRAO 140 (fig. \ref{fig:intNRAO140}) is a relativistic jet exhibiting high pattern speeds of $\beta_{app}\sim 13c$ and therefore $\theta_{ob}\leq 9^o$ \citep{Lister:2009}.  The intensity profile of NRAO 140 displays an asymmetry in the form of a hump in one tail of the profile consistent with our theoretical intensity profiles.

Further details regarding the observations of the intensity profiles and other observables covered in sections \S\ref{section:intensity}, \S\ref{section:RM}, and \S\ref{section:pi} are as follows: The observations for 3C 78 were carried out with the VLBA and Effelsberg on September 10, 2005 \citep{Kharb:2009}. NRAO 140 was observed as part of the MOJAVE program with the VLBA on August 9, 2007. The slices in total intensity (and fractional polarization, see figure \ref{fig:obs_PI}) were obtained with the task SLICE in AIPS. These were obtained roughly perpendicular to the local jet direction, and at a distance of approximately 5 mas for 3C 78, and 4.5 mas for NRAO 140.  We note that the observed profiles of total intensity (and fractional polarization) do change somewhat along the jet at different distances from the core. Therefore, to overcome this problem in future works, it will be necessary to study a statistically significant sample of jets and multiple slices within each jet.
\section{\normalsize Faraday Rotation Profiles}
\label{section:RM}
A linearly polarized electromagnetic wave can be decomposed into left-handed and right-handed circular polarization states which propagate through a magnetized ionic plasma with different phase speeds, causing a rotation of the plane of polarization.  This effect, known as Faraday rotation, can be calculated in the cold plasma approximation wherein the angle through which the EVPA is rotated is 
\begin{gather}
\Delta\tilde{\chi}=\frac{e^3}{2\pi m_e^2c^4}\lambda'^2 \int{n_T'B_{\|}'}ds',
\end{gather}
where $B_{\|}'$ is the component of the jet frame magnetic field parallel to the jet frame photon propagation vector, $n_T'$ is the jet frame number density of thermal electrons, and $\lambda'$ is the jet frame electromagnetic wavelength \citep{Burn:1966}.

The discovery of unambiguous $RM$ gradients first in 3C 273 \citep{Asada:2002,Zavala:2005} and then in other AGN jets \citep[e.g.,][]{Kharb:2009,Croke:2010} is consistent with the prediction that a large-scale helical field will give rise to a smoothly changing $RM$ across the jet \citep{Laing:1981,Blandford:1993}.  Various parsec scale polarimetric observations suggest the Faraday screen responsible for $RM$ gradients is not cospatial with the synchrotron emitting region \citep[e.g.,][]{Zavala:2004,Kharb:2009}.  The remaining probable locations of the Faraday screen are the sheath of the jet, the broad line region (BLR), and the narrow line region (NLR).  However, the BLR and NLR have been ruled out specifically in 3C 273 by the short time variability of $RM$ \citep{Asada:2008a} and in general by volume filling factor arguments and other polarimetric data \citep{Zavala:2004}.  Thus, in this paper, we model the Faraday screen as a cylindrical shell of thermal particles (a sheath) that likewise carries the helical field defined in equation (\ref{eqn:B}), but primarily resides outside of the synchrotron emitting spine of the jet.

The radial thermal electron density profile function we have chosen is
\begin{gather}
n_T'\propto\rho^6 \exp{\left(-\rho^2\right)},
\label{eqn:density}
\end{gather}
within the emission region where $\rho<1$ the thermal electron density goes as $n_T' \sim \rho^6$ and is therefore suppressed until $\rho \sim 1$ where it sharply rises.  This ensures the Faraday rotation occurs primarily in a cylindrical shell surrounding the jet up to a cylindrical radius of $\rho=1.6$ (see figure \ref{fig:field}).  Consequently, to match the observational evidence in our model there is little overlap between the emission region and the Faraday screen.  
\begin{figure}
\centering
\subfigure[][\label{fig:densities}]{\includegraphics[trim = 5mm 1mm 1mm 1mm, clip, width=2.3in]{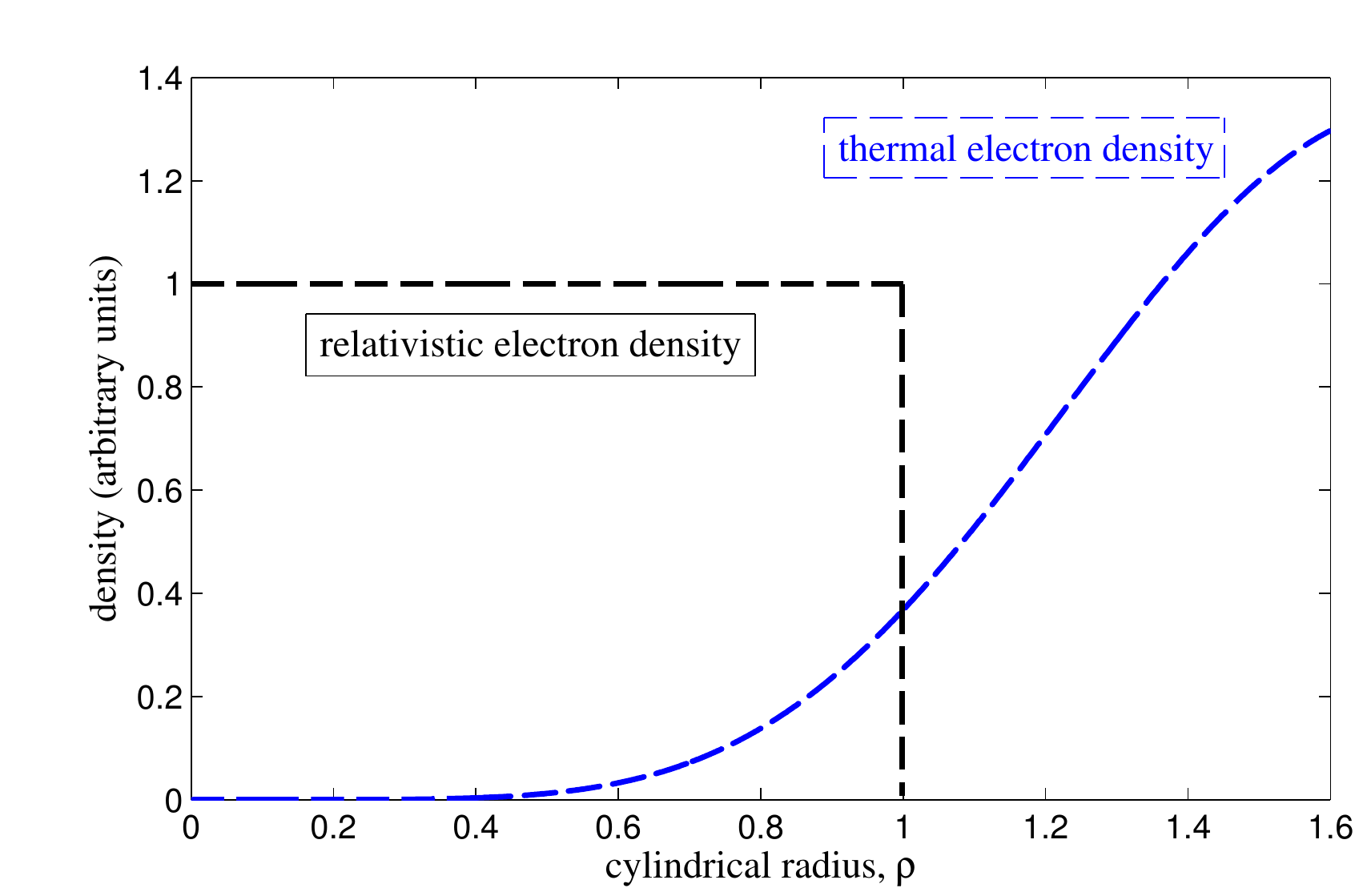}}
\subfigure[][\label{fig:theory_RM}]{\includegraphics[trim = 1mm 1mm 2mm 2mm, clip, width=2.3in]{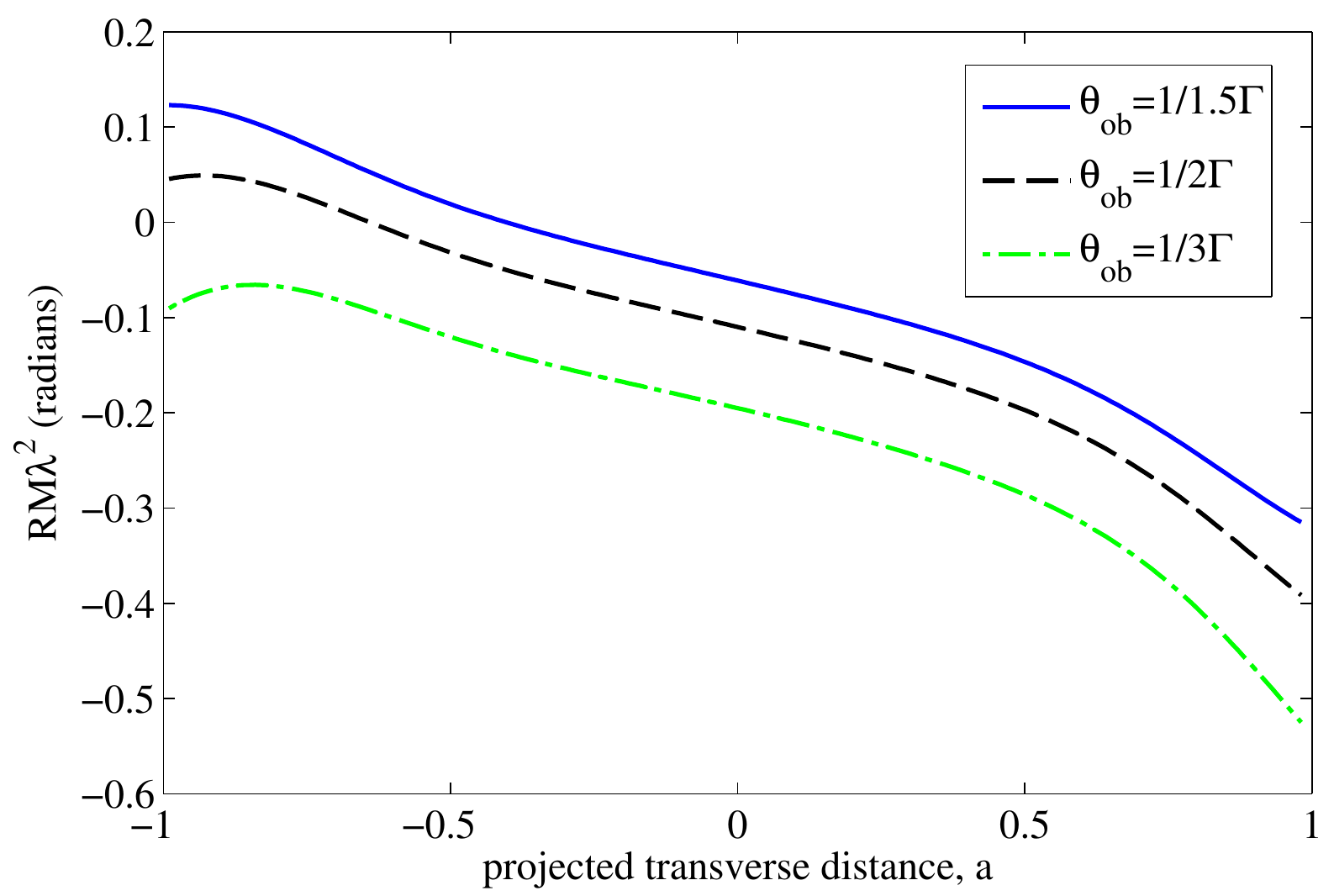}}
\subfigure[][\label{fig:obs_RM}]{\includegraphics[trim = 12mm 5mm 12mm 10mm, clip, width=2.3in]{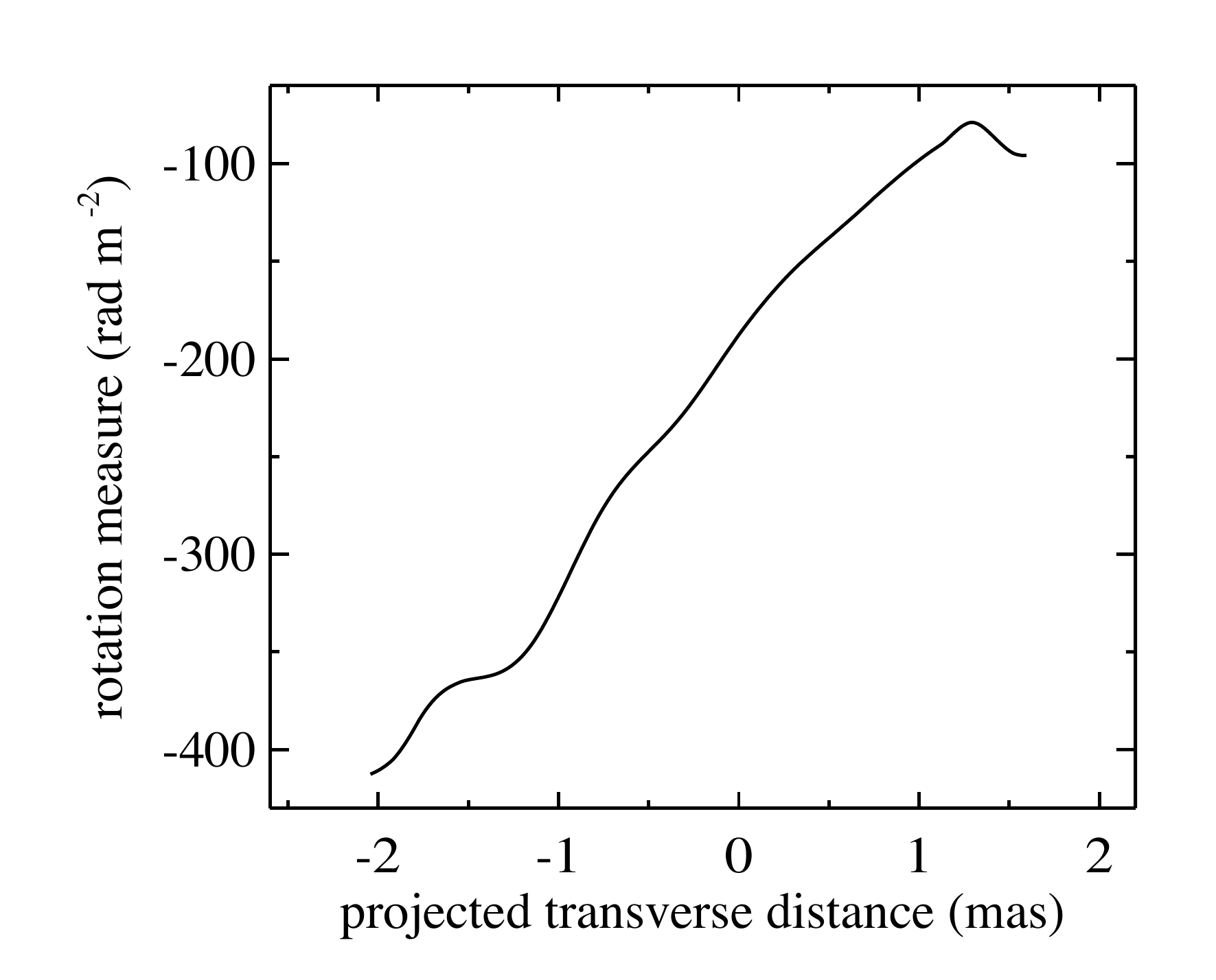}}
\caption{(a)  The dotted line displays the density profile of nonthermal synchrotron emitting particles.  The solid line shows equation (\ref{eqn:density}) which is the density profile of thermal particles in the Faraday screen.  The small amount of overlap between the two profiles produces little internal Faraday rotation.  (b)  Theoretical profiles of Faraday rotation $\Delta \tilde{\chi}_F$.  The primary effect viewing angle has on such profiles is to shift them up or down.  The amplitude $\left|RM\right|$ is higher on the right side of the jet because the magnetic field in the sheath (see fig. \ref{fig:field}) is more aligned with the line of sight on the right side of the jet than on the left. (c)  This $RM$ slice for 3C 78 was obtained where the $RM$ gradient was clearly observed, at a distance of $\sim 4$ mas from the core. Note that the $\left|RM\right|$ is higher on the left side.  This implies the sheath magnetic field is more aligned with the line of sight on the left side than on the right side of the jet.  The beam-size for this image is $2.5$ mas $\times$ 2.5 mas; for further error analysis, see \protect\cite{Kharb:2009}.} \label{fig:RMdetails}
\end{figure}
We carry out the calculation for the Faraday rotation in the rest frame jet assuming that the Faraday rotating sheath is moving at the same speed as the jet ($\Gamma=10$): 
\begin{gather}
\Delta \chi_F=A\int{n_T'\left|B'\right|\cos{\chi'}}ds'.
\label{eqn:RMs}
\end{gather}
Figure \ref{fig:theory_RM} shows the calculated invariant quantity $\Delta \tilde{\chi}_F=RM \lambda^2$ from equation (\ref{eqn:RMs}) for different viewing angles.  The normalization, $A$, is not as important as the shape of the $RM$ profile, but is set to give amplitudes of $\Delta \chi_F$ more or less consistent with observations.  As expected there is a gradient in $RM$ across the jet which is due to the line of sight component of toroidal magnetic field, $B_{\phi}$, changing across the jet.  The asymmetry in the amplitude of $RM$ in the jet is also apparent: the $\left|RM\right|$ on the left side of the jet is less than the $\left|RM\right|$ on the right side.  This is due to the axial field's contribution to the jet frame line of sight magnetic field.  For viewing angles $\theta_{ob} < 1/\Gamma$ the sheath magnetic field (see fig. \ref{fig:field}) is aligned with the line of sight on the right side more than on the left side of the jet.

The observed $RM$ profile for 3C 78 \citep{Kharb:2009} in figure \ref{fig:obs_RM} displays the asymmetry discussed above: the magnitude is greater on one side (the left) of the jet than it is on the other, suggesting the sheath magnetic field is more aligned with the line of sight on the left side of the jet.  Other observed $RM$ profiles reveal behavior more similar to our calculated profiles.  \cite{Asada:2008b}, for example, detect an $RM$ gradient using a different cut of NRAO 140 wherein the $RM$ changes sign across the jet with the gradient of $RM$ steepening toward the edges of the jet.  Also of note, \cite{Asada:2008b} finds that the magnitude of $RM$ is roughly symmetric across the jet, suggesting that the axial field is not contributing to the $RM$.  This symmetry in the magnitude of the $RM$ suggests that either the jet frame line of sight is orthogonal to the jet axis (i.e. $\theta_{ob}=1/\Gamma$) or the sheath field could be toroidally dominated.

\section{Fractional Polarization Profiles}
\label{section:pi}
The linear fractional polarization, $\Pi$, primarily depends on the geometry of the magnetic field in the emitting region.  If the magnetic field is isotropic and disordered in regions smaller than the beam, then $\Pi \sim 0$, and if the magnetic field is uniformly oriented, then the fractional polarization reaches a maximum of $\Pi=(p+1)/(p+7/3)\sim 0.7$ for typical values of the electron index, $p\sim2-3$ \citep{Pacholczyk:1970}.  Thus, lines of sight that pass through the edges of a jet with a helical field will encounter a more uniform field than lines of sight passing through the center of the jet, providing an explanation for the often observed rise in fractional polarization towards the edge of jets (though it should be noted that observations of fractional polarization at the edges of jets often have large errors).  In observational transverse profiles, the linear polarization tends to rise towards the edges of the jet, and $\Pi$ is often significantly higher on one edge of the jet \citep{Attridge:1999,Pushkarev:2005,Zavala:2005,Gomez:2008}.  This gradient in $\Pi$ across the jet (in which $\Pi$ is higher on one edge of the jet) can also be interpreted as a signature of a helical field (discussed below).  Observed polarization profiles also tend to be more strongly asymmetric than intensity profiles as seen, for example, by comparing figure \ref{fig:int_obs} to figure \ref{fig:obs_PI} below.

To calculate the fractional linear polarization for a synchrotron emitting plasma with relativistic bulk motion we follow the procedure outlined in \cite{Lyutikov:2003a,Lyutikov:2005} and calculate the Stokes parameters $I$, $Q$, and $U$ ($V=0$):
\begin{align}
	Q & \propto \int^{S}_{0}{ \left(B \sin{\chi'}\right)^{(p+1)/2} \cos{2\left(\tilde{\chi}+\Delta \tilde{\chi}_F\right)}}ds \notag \\
	U & \propto \int^{S}_{0}{\left(B \sin{\chi'}\right)^{(p+1)/2} \sin{2\left(\tilde{\chi}+\Delta \tilde{\chi}_F\right)}}ds,
	\label{eqn:stokes}
\end{align}
where $\tilde{\chi}$ is the angle the EVPA makes with the projection of the jet axis onto the sky (measured clockwise), and $\Delta \tilde{\chi}_F$ is the angle through which the EVPA is rotated in the foreground Faraday screen (see \S\ref{section:RM}).  If there were no bulk relativistic motion involved, the observed $\tilde{\chi}$ would be perpendicular to the projection of the jet magnetic field onto the sky.  However, in our case where the emitting fluid element has a relativistic bulk velocity we incorporate the effects of relativistic aberration on $\tilde{\chi}$   \citep{Blandford:1979,Lyutikov:2003a,Lyutikov:2005}.
\begin{figure}
\centering
\subfigure[][\label{fig:Pi_uncon}]{\includegraphics[width=3in]{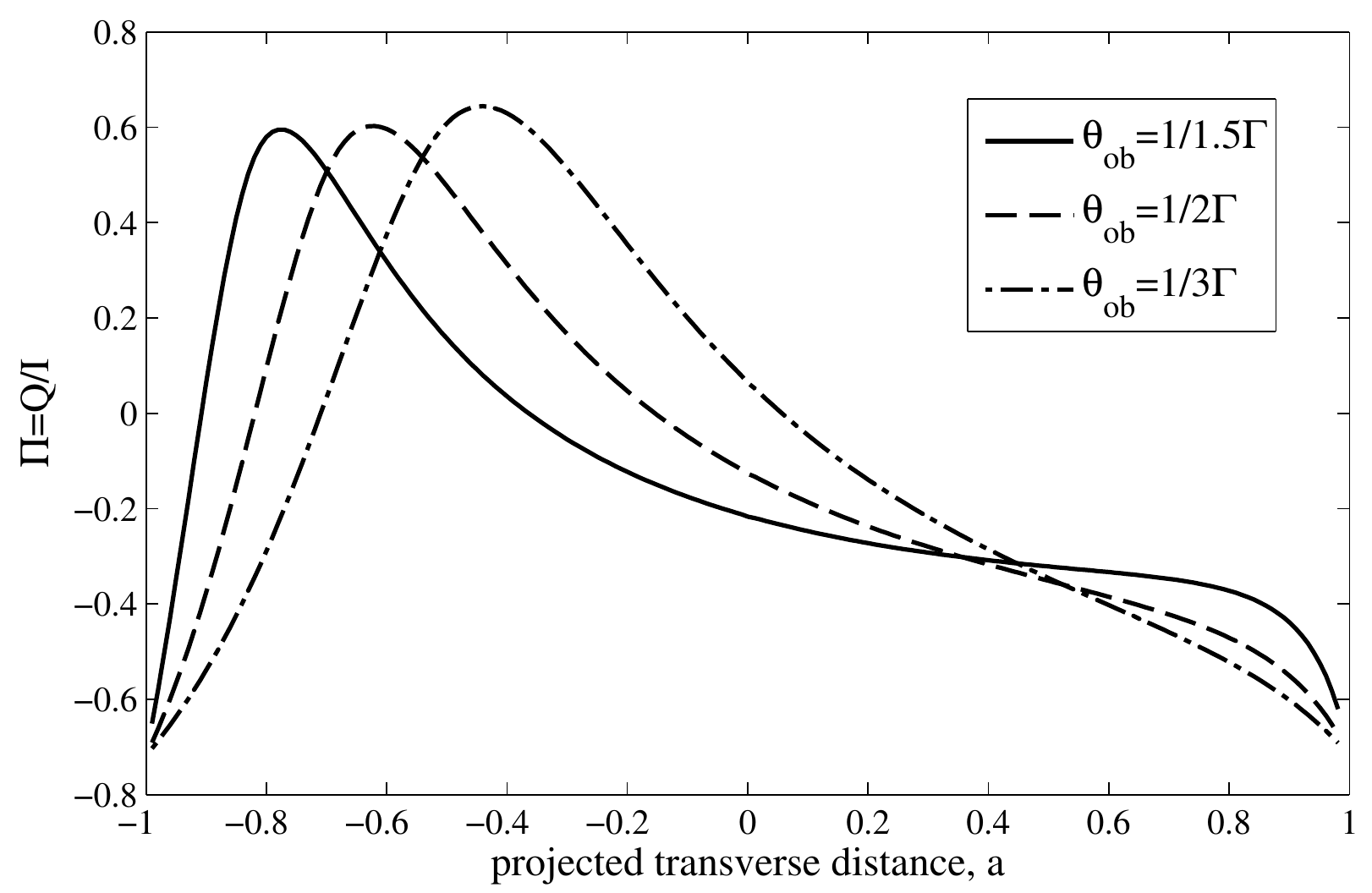}}\quad
\subfigure[ ][\label{fig:Pi_con}]{\includegraphics[width=3in]{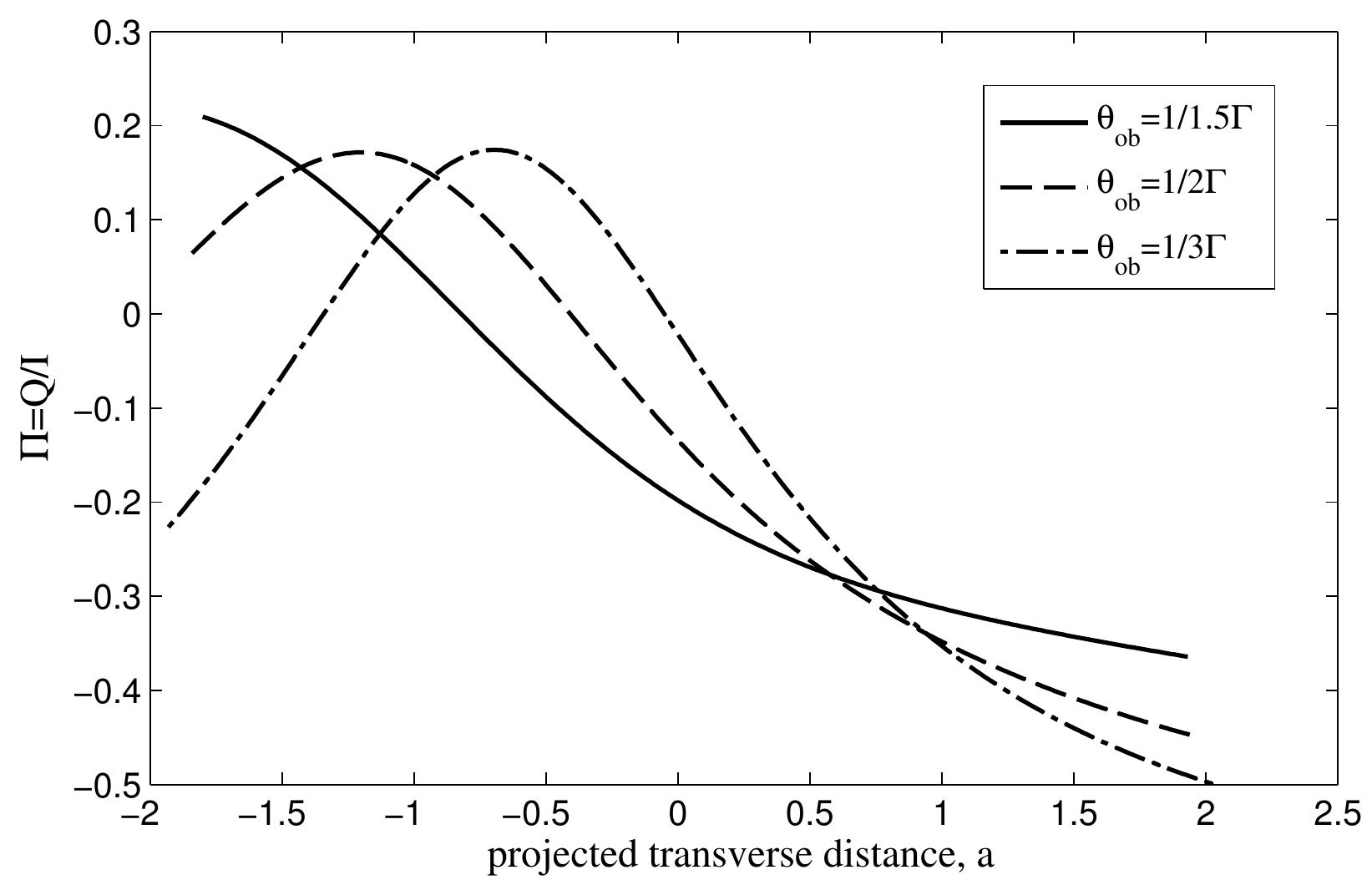}} \\
\caption{(a) Shown here are profiles of theoretical linear fractional polarization, $\Pi=Q/I$, for different viewing angles.  The sign of the polarization fraction refers to the EVPA direction: negative $\Pi$ corresponds to an EVPA that is perpendicular to the projected jet axis on the sky and positive $\Pi$ corresponds to an EVPA that is parallel to the projected jet axis on the sky.  As expected for large-scale magnetic fields, the polarization increases towards the edges of the jet.  The most notable qualitative asymmetrical feature is that the left side of the polarization profile has a much steeper gradient than the right side.  (b)  This plot shows the same theoretical profiles convolved with a Gaussian beam.  In these plots the primary effect is that the polarization of the left side of the jet is lower than the right side.} \label{fig:theory_pi}
\end{figure}
\begin{figure}
\centering
\subfigure[ ][\label{fig:PI3C78}]{\includegraphics[width=3.0685in]{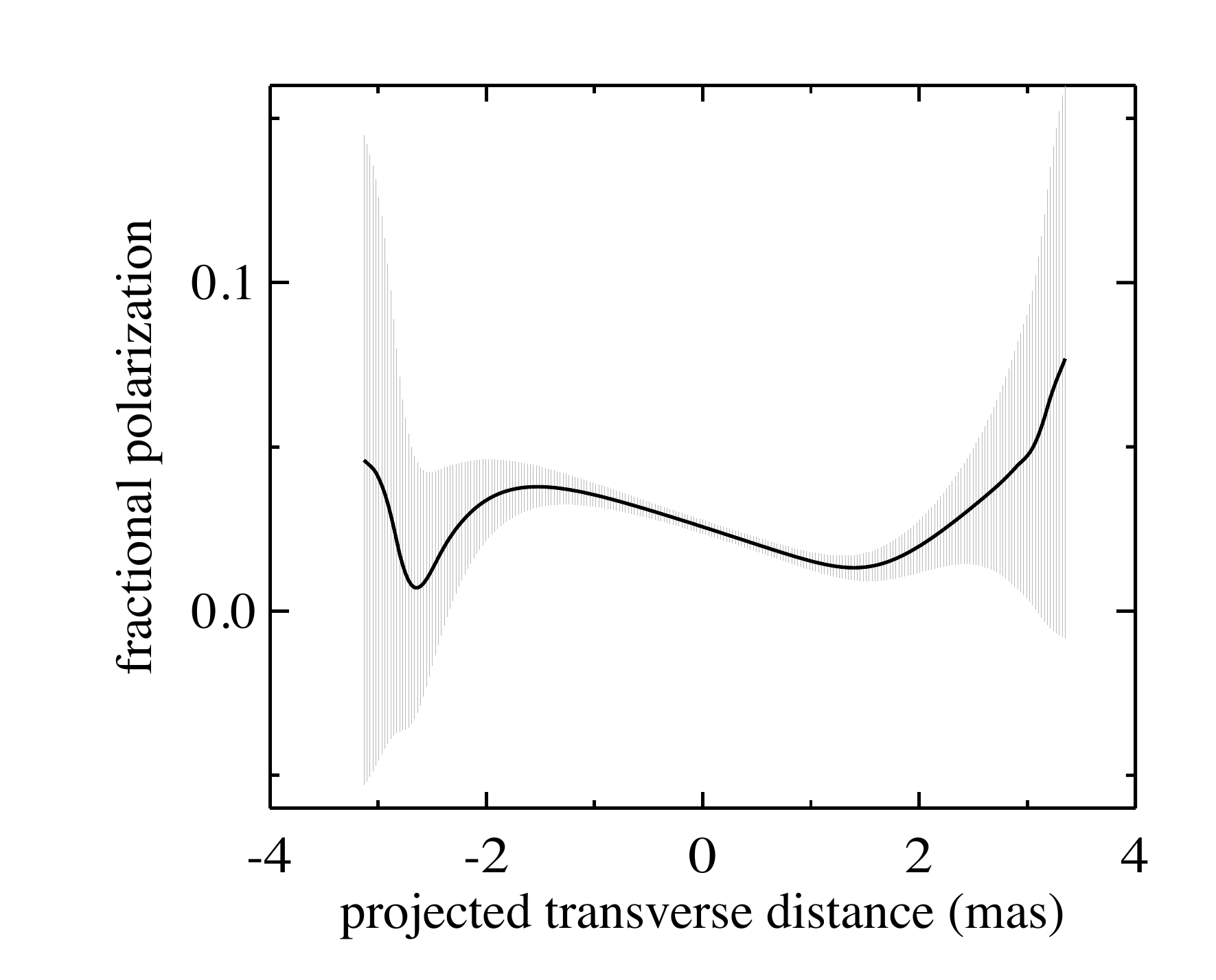}}
\subfigure[ ][\label{fig:PINRAO140}]{\includegraphics[width=3.0685in]{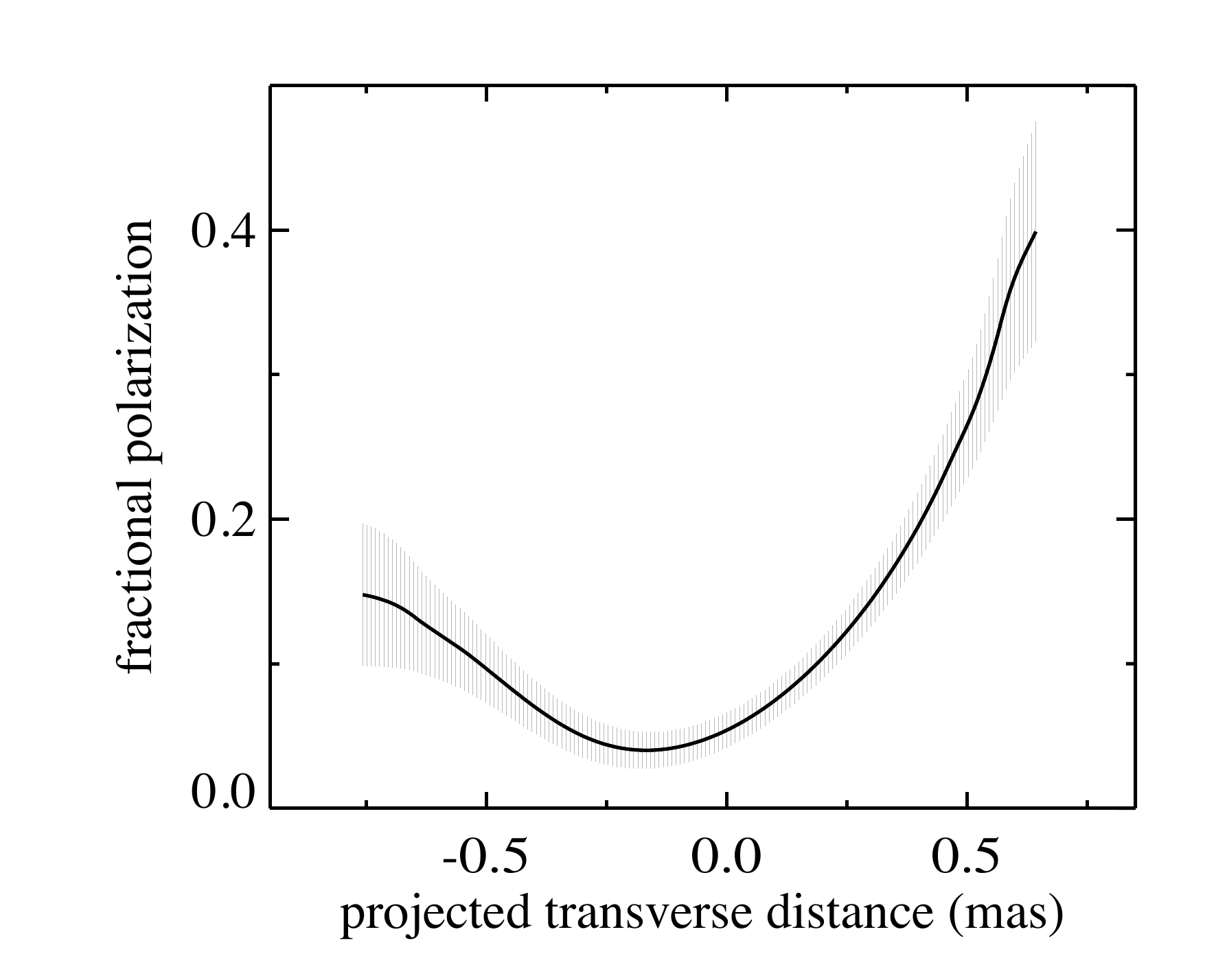}}
\caption{The slices in fractional polarization (and total intensity, see figure \ref{fig:int_obs}) were obtained with the task SLICE in AIPS. These were obtained roughly perpendicular to the local jet direction, and at a distance of approximately 5 mas for 3C 78, and 4 mas for NRAO 140.  The fractional polarization profiles also indicate the +/- sigma errors.  (a)  In this observed polarization profile of the parsec-scale jet of 3C 78, the large errors near the edge of the jet make it difficult to conclude much about the behavior (such as the dip on the left), except that the polarization appears to higher on the right side of the jet (the beam-size is $2.5$ mas $\times$ 2.5 mas).  (b)  In this observed profile of NRAO 140 there is no prominent ``bump", but the fractional polarization is clearly higher on the right side.  It also exhibits the expected increase in polarization towards the edges of the jet (the beam-size is $0.9$ mas $\times$ 0.6 mas).}
\label{fig:obs_PI}
\end{figure}
Equations (\ref{eqn:Int}) and (\ref{eqn:stokes}) allow us to calculate the fractional polarization:
\begin{equation}
\Pi=\frac{\sqrt{Q^2+U^2}}{I}.
\label{eqn:Pi}
\end{equation}
We found the $RM$ profile calculated in \S\ref{section:RM} had no effect on the qualitative features of the profiles.  For this reason we set $\Delta \tilde{\chi}_F=0$ in order to calculate the intrinsic EVPA angle which, given the cylindrical symmetry of the jet, can only be parallel or perpendicular to the jet axis on the sky \citep{Lyutikov:2005}.  Thus, we plot
\begin{equation}
\Pi=\frac{Q}{I},
\label{eqn:Pi2}
\end{equation}
which displays both the linear fractional polarization ($U$ is always zero due to cylindrical symmetry) and the EVPA direction.  When $\Pi$ is negative, the EVPA is perpendicular to the jet because $\tilde{\chi}=\pi/2$, or $Q\propto\cos{2\tilde{\chi}}<0$.  When $\Pi$ is positive then the EVPA is parallel to the jet, so $\tilde{\chi}=0$ and $Q\propto\cos{2\tilde{\chi}}>0$.

The calculated fractional polarization profiles are shown in unconvolved form in figure \ref{fig:Pi_uncon}.  The magnitude of linear polarization reaches the maximum at the edges of the jet, decreases towards the center, then, after reaching zero, increases again in the center. (The increase in $\Pi$ in the center region would appear as a ``bump" if the magnitude of $\Pi$ were plotted.)  This ``bump" corresponds to an EVPA flip from perpendicular to the jet axis on the edges to parallel to the jet in the middle region of the jet.  These flips have also been predicted by other helical models \citep[e.g.][]{Laing:1981,Lyutikov:2005,Broderick:2010}.  It is important to note that this ``bump" is not a robust feature of our calculations.  That is, for other configurations of the magnetic field such as the diffuse pinch, such a ``bump" does not appear for all viewing angles.  The most robust feature of the calculated profiles is the difference in the gradient of linear polarization across the jet.  The polarization decreases to zero on the left side of the profiles much more quickly than on the right side.

The profiles shown in figure \ref{fig:Pi_uncon} are unlikely to be directly detected because the sharp features will be smoothed by a finite beamwidth.  Thus we convolve each of the stokes parameters with a beam which has a standard deviation of $\sigma=0.2$ units (where the true projected jet diameter is $2$ units).  We define the edges of the convolved polarization profiles as the points at which the convolved intensity reaches 0.5\% of the intensity maximum.  These points represent the lower limit of the detector.  The results are shown in figure \ref{fig:Pi_con}.  The most robust feature is a gradient in polarization: the left side of the jet has a lower polarization than the right side.  As in the unconvolved case, EVPA flips occur as the linear polarization passes through zero.  The number of flips in our calculated profile depend on the intensity level at which we cut off the polarization profile. (For example, if the $\Pi$ profile were cut off at $5\%$ of the maximum intensity EVPA flips would be rarer.)  Convolving the $\Pi$ profiles also moves the ``bump" closer to the edge of the jet, and in the case of $\theta_{ob}=1/1.5\Gamma$ the ``bump" is not apparent at all and the EVPA flips only once or not at all going across the jet.  

The most robust prediction from our convolved theoretical profiles, a gradient in $\Pi$ across the jet, is seen in 3C 78 and NRAO 140 (fig. \ref{fig:obs_PI}) and elsewhere in the literature \citep{Attridge:1999,Pushkarev:2005,Zavala:2005,Gomez:2008}.  We note that the observed profiles of fractional polarization (and total intensity) do change slightly along the jet at different distances from the core.  Also, just as in the observed profiles of intensity (fig. \ref{fig:int_obs}) and polarization (fig. \ref{fig:obs_PI}), the calculated unconvolved and convolved polarization profiles display significantly more asymmetry than the (unconvolved) calculated intensity profiles.  However, our model does not explain the significantly lower polarization level of 3C 78 (0\% to 10\%) compared to most of our theoretical curves and NRAO 140.  This discrepancy can be due to a high degree of Faraday depolarization in the regions surrounding 3C 78's jet, or because of a significant disordered component to the jet magnetic field \citep{Cawthorne:1993,Zavala:2004}.

EVPA flips are not seen in 3C 78 and NRAO 140.  Two explanations for this absence are possible: (1) the magnetic field configuration in the emission region is different from that used in this work (e.g. a diffuse pinch field instead of a reverse pinch field), or (2) the EVPA flip occurs near the edge of the jet where the signal to noise ratio is too low for detection.  VLBI EVPA flips as predicted in the unconvolved profiles are detected in some jets (Kharb et al., in prep).
\section{\normalsize Spectral Index Profiles}
\label{section:alpha}
\begin{figure}
\centering
{\includegraphics[width=3in]{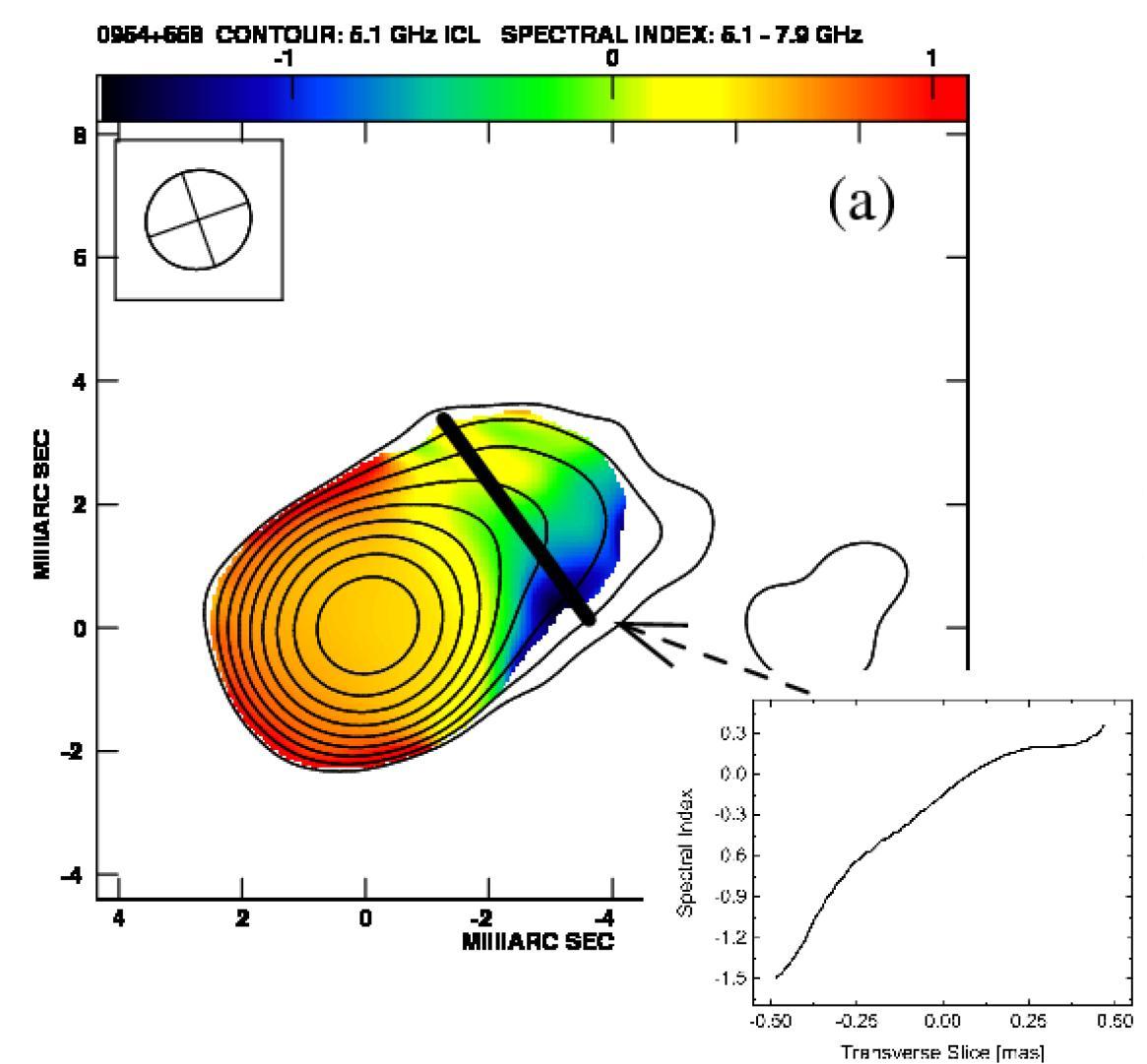}}
\caption{Observed spectral index map and profile for blazar 0954+658 (figure from \citealt{O'Sullivan:2009}).  Note the transition from a symmetric self-absorbed core to a region with a gradient as expected if the differential optical depth is inducing the spectral index gradient.  The spectral index gradient is seen just upstream from where the jet bends, thus this gradient may also be caused by the jet interacting with its surrounding medium.}
\label{fig:obs_spectral}
\end{figure}

Typical AGN jet morphology consists of an unresolved bright core with a flat or inverted spectral index, $\alpha \sim -1$ to $0$ ($I_{\nu} \propto \nu^{-\alpha}$), and a steep spectrum jet with a spectral index of $\alpha \sim 0.7$ \citep[e.g.][]{Zensus:1997}.  In some cases the transition from flat or inverted spectrum core to steep spectrum jet is thought to mark the transition between the compact optically thick portion of the jet and the optically thin portion of the jet \citep{Marscher:2009}.  Individual observed spectral index maps of AGN jets sometimes reveal more complicated behavior such as asymmetries in the tranverse direction across the jet as seen in figure \ref{fig:obs_spectral} \citep[e.g.,][]{Kharb:2009,O'Sullivan:2009, Savolainen:2008}.  These gradients in the spectral index are usually thought to indicate that the jet is interacting with an inhomogeneous external medium (jet-cloud interactions).  A large-scale helical magnetic field may also produce spectral index gradients.  We explore two possible ways helical fields could affect the spectral index: anisotropic particle distribution functions (\S\ref{section:p_effects}) and optical depth effects (\S\ref{section:tau_effects}).
\begin{figure}
\centering{
\subfigure[][\label{fig:spectral_schematic}]{\includegraphics[width=2.7in]{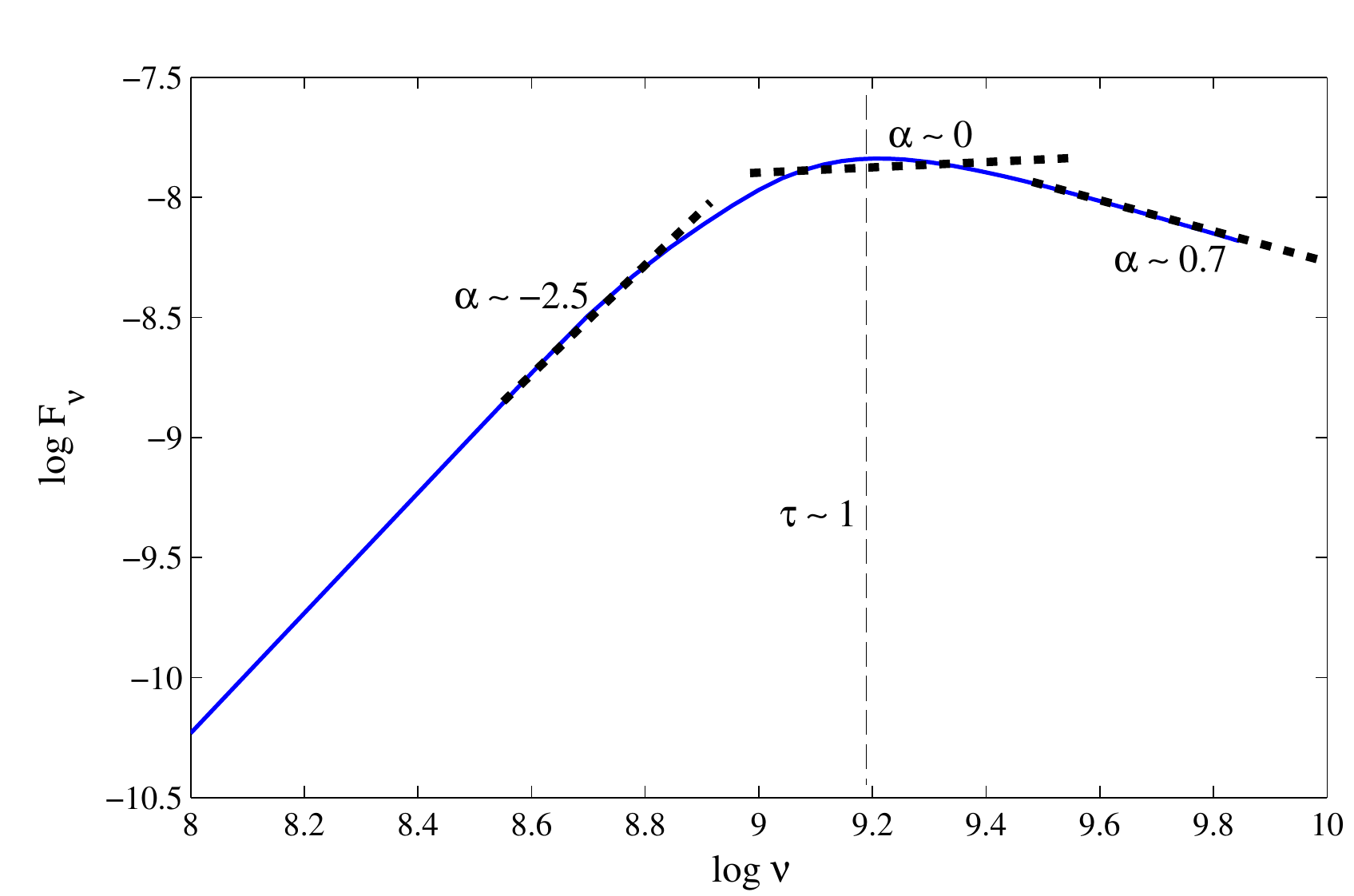}}
\subfigure[][\label{fig:spectral_tau}]{\includegraphics[width=2.7in]{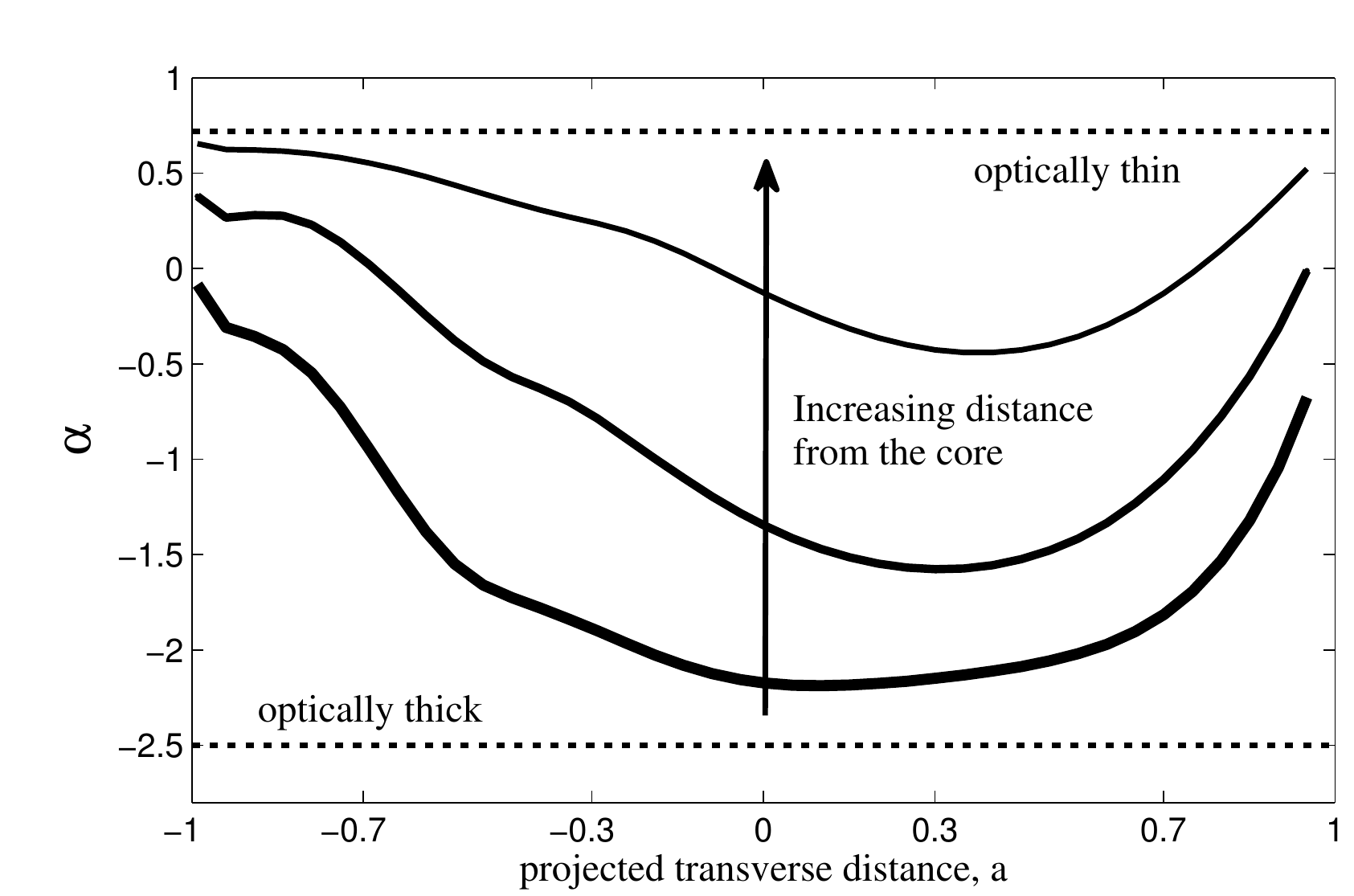}}
\subfigure[ ][\label{fig:spectral_p}]{\includegraphics[width=2.7in]{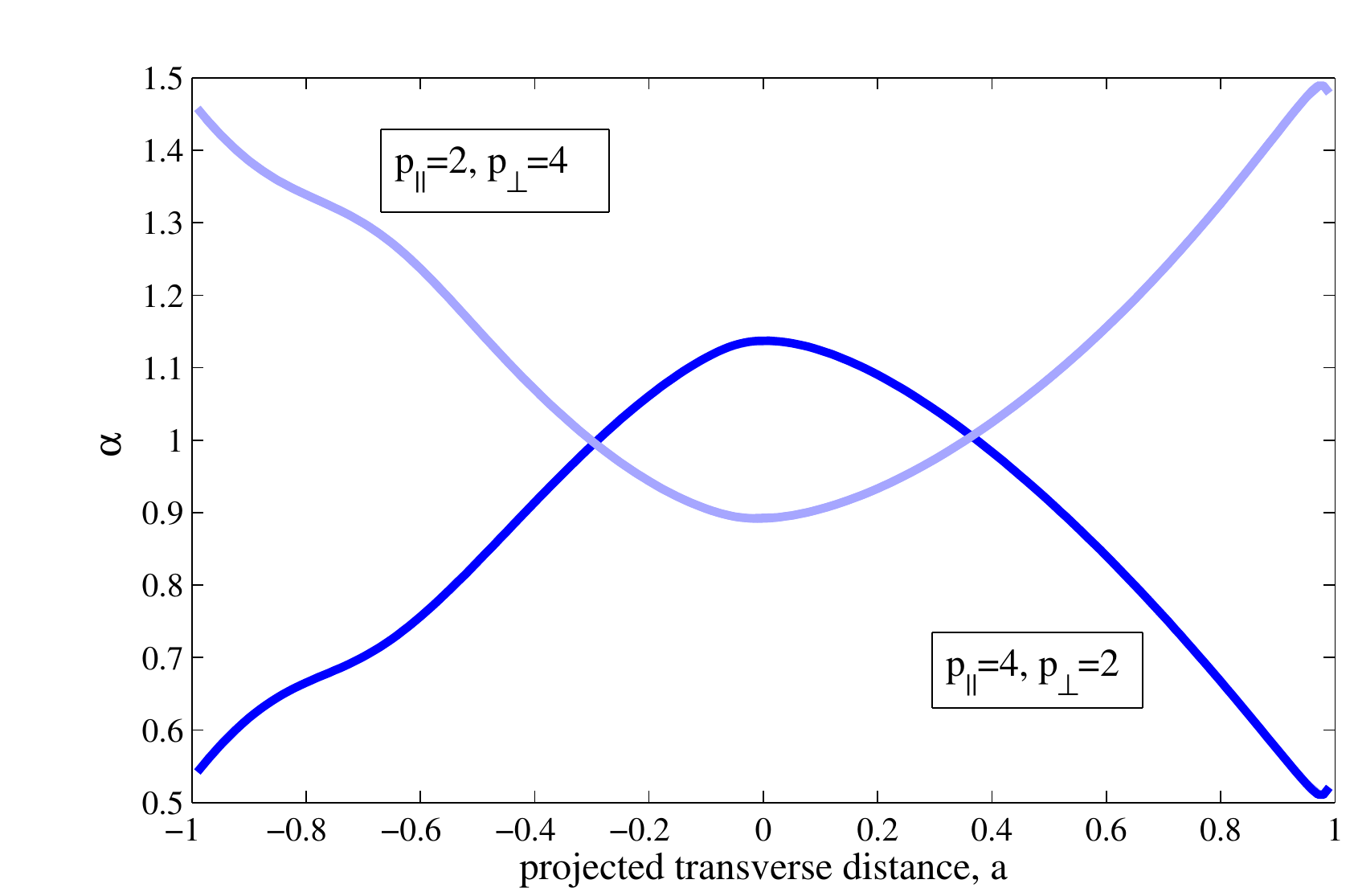}}}\\
\caption{(a) This figure uses a synchrotron spectrum from a homogeneous source with a power-law distribution to illustrate how optical depth effects can create a spectral index gradient.  The dotted lines represent power fits to different regions of the spectrum.  If the left side of the jet has a lower average optical depth than the right side of the jet then the left side of the jet will have $\alpha \sim 0.7$ range while the right side of the jet will be in the $\alpha \sim -2.5$ to $0$ range.  (b) Spectral index profiles at different distances from the supermassive black hole.  Note that such profiles have maximum changes of spectral index of $\Delta \alpha_{max} \sim 1$.  (c) Spectral index profiles from an anisotropic distribution.  Note that such profiles have maximum changes of $\Delta \alpha_{max} \sim 0.6$ and are not very asymmetrical.} \label{alpha}
\end{figure}
\subsection{\normalsize Anisotropic Distributions}
\label{section:p_effects}
In many shock particle acceleration models the electron distribution function is anisotropic \citep[e.g.,][]{Lloyd:2000}.  Particle in cell (PIC) simulations of shock acceleration occasionally find the electron index is dependent on the angle between the line of sight and the magnetic field, $\chi'$ (Spitkovsky, private communication).  We investigate the possibility that an anisotropy in the electron index $p$ is produced by the particle acceleration process or another unknown mechanism and is not isotropized too quickly by pitch-angle diffusion \citep[ch. 12,][]{Kulsrud:2005}.  If $p$ changes with the pitch angle of the relativistic electrons (the pitch angle is the angle between the electron's velocity vector and the magnetic field), then the spectral index $\alpha$ will depend on $\chi'$.  That is, for a given angle between the line of sight and the magnetic field, $\chi'$, only the population of electrons whose pitch angle $\alpha=\chi'$ will be observed due to the relativistic beaming of individual electrons.  Therefore, the observed spectral index will be $\alpha=(p(\chi')-1)/2$ in an optically thin jet.  

The distribution we investigate changes linearly from one value ($p_{\perp}$) when $\chi'=\pi/2$ to another ($p_{\|}$) when $\chi'=0$:
\begin{align}
dn &=K_e \gamma^{-p(\chi')}d\gamma \notag \\
p(\chi') &= \frac{2p_{\perp}\chi'}{\pi}+p_{\|}\left(1-\frac{2 \chi'}{\pi}\right) \notag \\
\alpha &= \frac{p(\chi')-1}{2}.
\label{eqn:anisotropic}
\end{align}
We obtain the spectral index profile in figure \ref{fig:spectral_p} by assuming an optically thin jet and integrating equation (\ref{eqn:Int}) with $p\rightarrow p(\chi')$.  The spectral index is constructed by numerically evaluating equation (\ref{eqn:Int}) for different $\nu$, thereby constructing an optically-thin power-law spectrum.  The range of values used for $p_{\|}$ and $p_{\perp}$ is the strong shock value of the electron index $p=2$ \citep{Drury:1983} and a steeper spectrum of $p=4$ consonant with a radiatively cooled spectrum ($p\rightarrow p+1$) with an injected electron index of $p_{inj}=3$ seen in the cosmic rays.  The results of these calculations are discussed in \S\ref{section:spec_results}.
\subsection{\normalsize Optical Depth Effects}
\label{section:tau_effects}  
We assume AGN jets consist of a compact optically thick core which gradually evolves into an optically thin jet.  The existence of helical fields will modify this transition so that one side of the jet will become optically thin at a different core distance, $z$, than the other side of the jet.  Therefore, there will be a region in the jet where one side has a different optical depth than the other side of the jet.  The average synchrotron optical depth, $\tau$ is defined as \citep{Rybicki:1979}:
\begin{gather}
\tau_{\nu} \propto \int^{S}_{0}{\left(B'\sin \chi' \right)^{(p+2)/2}\nu^{-(p+4)/2}ds}.
\label{eqn:tau}
\end{gather}
Equation (\ref{eqn:tau}) reveals that the optical depth is an increasing function of $\sin{\chi'}$, implying that $\tau$ is higher on the right side of the jet where $\sin{\chi'} \sim 1$ than on the left side where $\sin{\chi'} \ll 1$ (see fig. \ref{fig:field}).  Thus, for a magnetic field configuration as viewed in figure \ref{fig:helix} the optical depth on the right side of the jet ($a>0$) will always be greater than the optical depth on the left side of the jet ($a<0$).

To gain insight into how the synchrotron optical depth changes from optically thick to optically thin, we estimate how the following quantities scale with $z$: the electron density, the magnetic field, and the length of the line sight passing through the jet.  While our jet is modeled as a cylinder, we derive the scalings as if the jet were conical.  Conical expansion of the jet flow implies that the electron density scales as $\propto z^{-2}$, and this expansion along with flux freezing implies the toroidal magnetic fields scale as $\propto z^{-1}$. (If the magnetic field were primarily poloidal then $B \propto z^{-2}$; however this difference does not appreciably change our results.)  The conical geometry implies that the line of sight distance scales as $\propto z$.  These scalings, along with equation (\ref{eqn:tau}) imply that
\begin{align}
\kappa \propto z^{-\frac{p+4}{2}} \notag \\
j\propto z^{-\frac{p+3}{2}},
\label{eqn:coeffs}
\end{align}
where $\kappa$ is the average absorption coefficient, $j$ is the emission function, and $p$ ($=2.4$) is the electron index.  To calculate the spectrum we numerically solve the rest frame transfer equations (equation (3.66) of \citealt{Pacholczyk:1970}) for the Stokes parameters which include the polarized emission and absorption coefficients that depend on $z$ as defined in equation (\ref{eqn:coeffs}).  After evaluating the radiative transfer equations and obtaining the total intensity numerically for different $\nu$, a spectrum is constructed over a one order of magnitude interval in frequency (e.g. $1-10$ GHz).  The spectral index is calculated by finding the best-fit linear slope of the log-log spectral interval using a reduced squares fit as illustrated in figure \ref{fig:spectral_schematic}.
\subsection{Spectral Index Results}
\label{section:spec_results}
The calculated profiles are shown in figures \ref{fig:spectral_tau} and \ref{fig:spectral_p}.  One notable difference between these figures is that the profile resulting from an anisotropic distribution function is not as asymmetric as that arising from optical depth effects, though this difference may disappear with a different form of equation (\ref{eqn:anisotropic}).  Two more general differences between optical depth effects and anisotropic distributions are: (1) optical depth effects produce a greater range in spectral index which include both inverted and steep spectral indices (theoretically, $\alpha=-2.5$ to $\sim0.7$), as opposed to anisotropic distribution functions which only produce steep spectral indices (theoretically, $\alpha=(p-1)/2=0.5$ to $1.5$), and (2) anisotropic distribution functions can produce spectral index gradients anywhere in the optically thin jet while optical depth effects give rise to such gradients only near the AGN core/optically thin jet boundary.  

We include an observed spectral index gradient in 0954+658 \citep{O'Sullivan:2009} which displays the expected behavior of a spectral index gradient caused by optical depth effects: the core is symmetrical and self-absorbed, and a gradient in spectral index is seen in the transition region between optically thick and optically thin.  Alternatively, because the spectral index gradient occurs just upstream from where the jet bends in 0954+658, the spectral index gradient may be the result of a jet-cloud interaction.  A spectral index gradient observed in 3C 273 \citep[fig. 1,][]{Savolainen:2008} is also consistent with a helical field inducing the gradient via optical depth effects.  The spectra associated with various components of the parsec scale jet of 3C 273 reveal a self-absorbed spectrum from the core to where the jet widens, at which point the south side of the jet (component B3) displays a self-absorbed spectrum while the north side (component B2) exhibits an optically thin spectrum.

Unfortunately, observations of spectral index gradients alone are explained equally well by jet-cloud interactions. However, as will be discussed in \S\ref{section:conclusion}, if such spectral index gradients correlate with asymmetrical features of other VLBI observables, the case for helical field induced spectral index gradients is strengthened.
\section{Discussion And Conclusions}
\label{section:conclusion}
We have shown that a cylindrical jet with an axially symmetric large-scale helical magnetic field produces asymmetrical transverse profiles for intensity, $RM$, fractional polarization, and spectral index.  The profiles' asymmetries arise from changes in $\left|\sin{\chi'}\right|$ (or $\cos{\chi'}$ for $RM$) across the jet, where $\chi'$ is the angle between the line of sight and magnetic field in the jet frame.  Therefore, the degree of asymmetry in these profiles depends on the magnetic field structure (assumed to be force-free in this work), the jet viewing angle, and the bulk Lorentz factor.  

Unknown structural details of the jet--including boundary conditions, certain aspects of the functional form of the magnetic field, and the relativistic particle density profile--introduce significant uncertainty into all of our calculated transverse profiles.  However, the qualitative features of the magnetic field used in this work (i.e. $B_{\phi}=0$ and $B_{z}=$ maximum on the jet axis and $\left|B\right|$ decreasing with cylindrical radius) are shared with a variety of analytic \citep[e.g.][]{Choudhuri:1986,Lynden-Bell:1996} and numerical \citep[e.g.,][]{Li:2006,McKinney:2006} models of AGN jets.  To probe whether the asymmetrical features highlighted here are dependent on the specific structural form of the jet assumed in this paper, we have calculated transverse profiles (not shown here) using alternate boundary conditions, magnetic field structures, and density profiles.  A force-free diffuse pinch field is tested, as well as several relativistic density profiles: $n_{rel}' \propto$ current$'^2$ (ohmic dissipation), $n_{rel}'\propto B'^2$ (equipartition), and a Gaussian density profile.  These alternatives and their permutations affect specific qualities of the profiles (discussed below), but they do not affect the robust asymmetrical features underlined in this work.

Another significant unknown not treated in this work is whether a disordered field component would erase the predicted asymmetries.  Regarding polarization, for example, a disordered field component decreases the fractional polarization of a uniform field by the factor $\sim B_0^2/(B_0^2+B_d^2)$, where $B_0$ is the uniform field and $B_d$ is the disordered field \citep{Burn:1966}.  For intensity profiles, increasing $B_d^2/B_0^2$ gradually erases the asymmetry because the disordered field component provides the same contribution to the synchrotron emission function on both sides of the jet.  Consequently, as $B_d^2/B_0^2$ is increased, the predicted asymmetries in this work would be diluted, though the exact ratio of $B_d^2/B_0^2$ at which the asymmetries become undetectable depends on how well the jet is resolved and the particular magnetic field structure of the jet.

Although the aforementioned unknowns in AGN jet structure are significant, some conclusions can be made about the theoretical profiles calculated in this paper.  For polarization profiles (\S\ref{section:pi}), the details of the jet structure unfortunately dictate how many EVPA flips occur across the jet and whether the polarization reaches zero.  However, the significant gradient in polarization for all of the various jet structures discussed above (except for those with a significant disordered field component) is a robust feature of our theoretical polarization profiles. For intensity profiles (\S\ref{section:intensity}) and spectral index profiles (\S\ref{section:alpha}), the qualitative features do not depend on the details of the jet structure.  Intensity profiles display a skewness that primarily depends on viewing angle and become bimodal for large or small viewing angles, $\theta_{ob} \gg or \ll 1/\Gamma$.  These bimodal intensity profiles may explain the observed edge-brightening in radio jets like M87 \citep{Reid:1989}.  Regarding spectral index profiles, there are two mechanisms by which a helical field gives rise to spectral gradients: optical depth effects and anisotropic particle distribution functions.  Optical depth effects produce spectral index gradients that range from inverted to steep spectral indices only on the boundary between the AGN core and the optically-thin jet.  Anisotropic particle distribution functions can induce gradients anywhere in the optically thin jet but have a more limited range of spectral indices, all of which are steep spectra.  For Faraday $RM$ profiles (\S\ref{section:RM}), many qualitative features do depend on specific, unknown features of the jet structure such as the presence of magnetic field reversals in the jet sheath and the functional form of the thermal particle density profile.  The implications of this uncertainty in $RM$ profiles is discussed below.

As we have shown, a helical magnetic field can clearly produce emission patterns that are asymmetrical, however, other phenomena such as jet-cloud interactions or a curved jet can as well.  Consequently, the ubiquity of asymmetrical profiles in parsec scale AGN jets alone lends little credence to the hypothesis that large-scale helical magnetic fields are present.  An unambiguous signature of large-scale helical magnetic fields would be asymmetrical features in profiles of different VLBI observables that correlate with one another as specified in this work.  The most straightforward correlation is between intensity profiles and linear polarization profiles:
\begin{itemize}
	\item Intensity profiles will be skewed such that the longer tail will be on the same side of the jet where the polarization is lower.
\end{itemize}
Detecting skewed intensity profiles may be difficult since the predicted asymmetry will be significantly smoothed out by the beam.  However, asymmetries in polarization profiles are easier to detect with a finite beam since the resulting profile will display a gradient in polarization.

Comparing intensity and polarization profiles with spectral index profiles is not as straightforward.  Ideally, all VLBI profiles should come from the same transverse cut on the jet, but care must be taken where spectral index gradients are present.  The differing optical depths in a transverse cut with a spectral index gradient will induce asymmetrical features in polarization and intensity profiles that may be difficult to disentangle from the asymmetrical features induced by helical fields.  Thus, it may be necessary to examine polarization and intensity profiles in optically thin regions downstream of observed spectral index gradients.  Assuming that (a) the same helical field is present at both the spectral index gradient cut as well as the intensity and polarization transverse cut, and (b) differing optical depth effects (\S\ref{section:tau_effects}) are responsible for the spectral index gradient, then:
\begin{itemize}
\item The side of the jet where the spectrum is more optically thin (or steeper) will be the side where the intensity profile has a long tail and where the polarization is lower.
\end{itemize}
However, if the particle distribution function of the relativistic electrons is anisotropic (\S\ref{section:p_effects}), there will be a correlation between the direction of the spectral index gradient and other VLBI observables, but we cannot determine what kind of correlation should exist without a better understanding of particle acceleration mechanisms in parsec scale AGN jets.  Correlations seen between spectral index gradients and other observables far away from the core in the optically thin region of the jet would support the anisotropic electron distribution function explanation.

$RM$ profiles probe the surrounding jet sheath, not the jet spine, where synchrotron emission takes place.  Unfortunately, the unknown nature of the magnetic field and thermal particles in the sheath make it difficult to identify robust features of $RM$ profiles besides the existence of transverse gradients.   Therefore, $RM$ profiles may not correlate with profiles of intensity, polarization, or spectral index.  However, if the spine and sheath of the jet contain ascending and descending magnetic flux as in magnetic tower models \citep[e.g.,][]{Lynden-Bell:1996} there may be a correlation between observables that probe the spine and $RM$, which probes the sheath.  Thus, comparing $RM$ gradients with other VLBI observables may further illuminate a possible connection in the magnetic fields between the jet spine and sheath.

In this work we have compared observed profiles for NRAO 140 and 3C 78 with our theoretical predictions.  For NRAO 140 the observed intensity profile (fig. \ref{fig:intNRAO140}) and the polarization profile (fig. \ref{fig:PINRAO140}) show the correlation expected from helical fields: the long tail of the intensity profile is on the side of the jet (the left) where the polarization fraction is low.  3C 78 is more ambiguous.  Its intensity profile (fig. \ref{fig:int3C78}) is not significantly skewed, but the polarization profile (fig. \ref{fig:PI3C78}) does exhibit a gradient with higher polarization on the right side of the jet, though even this is unclear due to the large errors on the jet's edges.  The spectral index map of 3C 78 \citep[not reproduced here, see fig. 4 of][]{Kharb:2009} is difficult to interpret as it reveals gradients in both directions in different parts of the jet well beyond the optically thick core.  If the polarization profile of 3C 78 is due to helical fields, a comparison can be made with its $RM$ profile shown in figure \ref{fig:obs_RM}.  The left side of the 3C 78's jet exhibits a higher $\left|RM\right|$, suggesting that 3C 78's magnetic field in the spine (i.e. emission region) and sheath is different from the assumed magnetic field structure assumed in this work (fig. \ref{fig:field}).  That is, the lines of sight through the left side of 3C 78's jet spine and sheath are more closely aligned with the magnetic field than those on the right side.  

To test whether the signature of large-scale helical fields exist, a systematic search for the correlations between different VLBI profiles described here is required.  This not only involves investigating as many resolved AGN jets as possible, but also involves examining multiple profiles at different distances from the core for a single jet.  In the future we intend to carry out such a search in the MOJAVE sample of radio loud jets with a focus on well-resolved jets.
\section*{Acknowledgments}
We thank D. Gabuzda, M. Lister, S. O'Sullivan, T. Hovatta, T. Savolainen, and the reviewer, T. Cawthorne, for their valuable comments on this work.  P.K. would like to acknowledge the significant contribution made by D. Gabuzda and P. Shastri in acquiring and analyzing the VLBI polarimetry data on 3C 78.  This research has made use of data from the MOJAVE database that is maintained by the MOJAVE team (Lister et al., 2009, AJ, 137, 3718).  E.C. gratefully acknowledges financial support from NASA \textit{Fermi} grant NNX10AO46G. 

\label{lastpage}
\end{document}